# Effect of independent parameters on nanoparticle sizes in magnetron-sputtering inert-gas condensation

Yizhou Wang[a,b], Evropi Toulkeridou[c]*, Abisegapriyan K.S.[a], Yair Ein-Eli[b,d,e], Panagiotis Grammatikopoulos[a,f,g]*

[a]Materials Science and Engineering, Guangdong Technion – Israel Institute of Technology, Shantou, Guangdong 515063, China

[b]Department of Materials Science and Engineering, Technion – Israel Institute of Technology, Haifa 3200003, Israel

[c]Escuela Técnica Superior de Ingeniería Industrial, Universidad de Castilla-La Mancha, 13071 Ciudad Real, Spain

[d]Grand Technion Energy Program (GTEP), Technion – Israel Institute of Technology, Haifa 3200003, Israel

[e]Israel National Institute of Energy Storage (INIES), Technion – Israel Institute of Technology, Haifa 3200003, Israel

[f]Guangdong Provincial Key Laboratory of Materials and Technologies for Energy Conversion, Guangdong Technion – Israel Institute of Technology, Shantou, Guangdong 515063, China

[g]Instituto Regional de Investigación Científica Aplicada (IRICA) and Departamento de Física, Universidad de Castilla-La Mancha, 13071 Ciudad Real, Spain

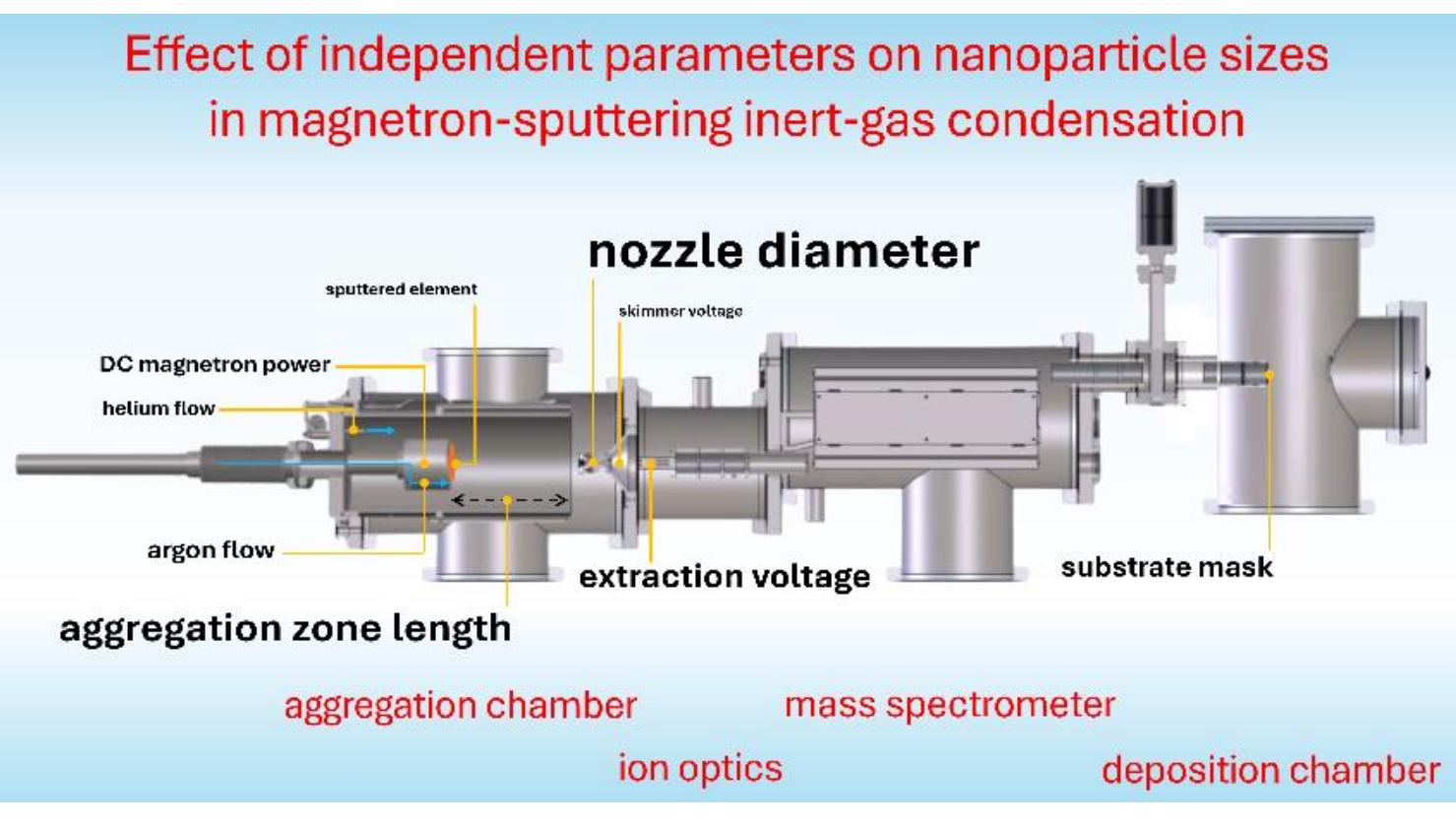

**ABSTRACT:** Magnetron-sputtering inert-gas condensation (MS-IGC) provides a scalable, environmentally friendly vapour-phase synthesis approach for preparing customised nanoparticles (NPs) with bespoke properties *via* fine-tuning several deposition parameters. However, this high-level control comes with a caveat: the synthesis mechanisms are affected by deposition parameters in complicated ways, often yielding unpredictable outputs. This report details the working mechanism of a typical MS-IGC system, achieving *in situ* synthesis, size screening, and directional deposition of nanoclusters through the synergistic operation of the three vacuum chambers (condensation, screening, and deposition). The study systematically explores the regulation laws of multiple key process parameters (e.g., inert-gas flows, aggregation length, etc.) on the formation, size distribution, and deposition behaviour of nanoclusters, to rationalise their chosen values toward optimised output. To this end, multiple linear regression analysis was performed to isolate the effect of each deposition parameter and, thus, quantify its effect on the NP size and size distribution. Our results indicate that parameters that may prolong the nascent NPs' residence inside the condensation chamber (most prominently, the exit nozzle diameter) can positively affect the NP final sizes. This study expands the understanding on NPs formation, enabling an optimised experimental control.

## 1. Introduction

Cluster beam deposition (CBD) is a relatively new solvent- and effluent-free technique to fabricate nanoparticles (NPs) with bespoke features often unattainable by wet chemistry growth methods [1-3]. The primary distinction among various CBD sources lies in the physical mechanism by which material is ejected from an initial bulk source. Common ejection methods include laser ablation, sputtering, thermal evaporation, and pulsed arc discharge. The design characteristics and applicable scenarios of different cluster sources have been systematically discussed in Chapter 3 of Ref. [1].

This study focuses on NP fabrication by magnetron-sputtering inert-gas condensation (MS-IGC) [4-6]. MS-IGC is a physical vapour deposition technique in which atoms are ejected from a solid target via magnetron sputtering and subsequently cooled and nucleated into clusters within a relatively high-pressure inert-gas atmosphere before being deposited onto a substrate. In this approach, a magnetically confined plasma (typically sustained in argon), enhances ionisation efficiency and sputtering yield, while the inert gas promotes rapid thermalisation of the sputtered species, enabling controlled nucleation, growth, and aggregation in the gas phase. The resulting NP size, crystallinity, and morphology are governed by key input parameters such as sputtering power, inert-gas pressure and composition, aggregation length, and gas flow rate, which collectively determine the balance between atom flux, cooling rate, and residence time in the condensation zone [7]. Owing to its clean, ligand-free synthesis environment and precise tunability, MS-IGC provides a versatile platform for systematically investigating how process parameters influence NP formation dynamics and functional properties. The growing emphasis on ecological design and manufacturing, together with inherent advantages such as sample purity, is actively driving its adoption in industrial-scale applications. [8,9].

MS-IGC is often praised for presenting a number of advantages such as pure cluster-support interfaces, NP soft landing, surface coverage control, mixing of various elements (if desired), and C-MOS compatibility, to name but a few. However, it is still hindered by the complex interplay between various deposition parameters, which, while offering unique possibilities for enhanced structural control, poses challenges in the determination of optimal design practices that may yield consistently reliable and uniform products [10]. To this day, various studies, both experimental and theoretical, have focused on elucidating this interplay and providing a roadmap for the consistent production of tailored nanoparticulated samples [11-15]. Though useful and intuitive, such studies typically provide qualitative design protocols, with only a few studies attempting to quantify their arguments (e.g., [16-20]), or recommend alternative strategies to increase NP sizes [21]. This work aspires to offer a quantitative analysis based on multiple linear regression, a statistical method used to quantify the relationship between a dependent variable and several independent variables simultaneously. Here, it is employed to evaluate how key MS-IGC process parameters, such as sputtering power, inert-gas flow, and aggregation length, collectively influence NP size. By fitting a linear model to the experimental data, the approach estimates the relative contribution and statistical significance of each parameter while accounting for their combined effects, thereby providing a systematic and transparent framework for identifying the dominant factors governing NP formation.

## 2. Methodology

### 2.1. Nanoparticle Fabrication and Characterisation

This work employed the ClusterGEN Max-100 third-generation cluster deposition system developed by Teer Coatings Co., Ltd. The system consists of three interconnected vacuum chambers (as shown in Figure 1), each of which performs a specific function during the preparation of NPs. The first chamber serves as the main aggregation zone, responsible for the nucleation and growth of NPs. The second chamber is equipped with a precision mass filter/spectrometer capable of precisely screening NPs based on their number of atoms (after the cluster beam was collimated via ion optics). The third chamber is designed for sample deposition. A substrate mask (often called a shadow or stencil mask) is a physical barrier placed near the substrate, exploiting the highly directional, collimated nature of the cluster beam to pattern structured nanoparticulate films. Aggregation chamber cooling is achieved using water or liquid nitrogen flowing through the double-walled structure of the magnetron head and a cooling jacket enveloping the chamber.

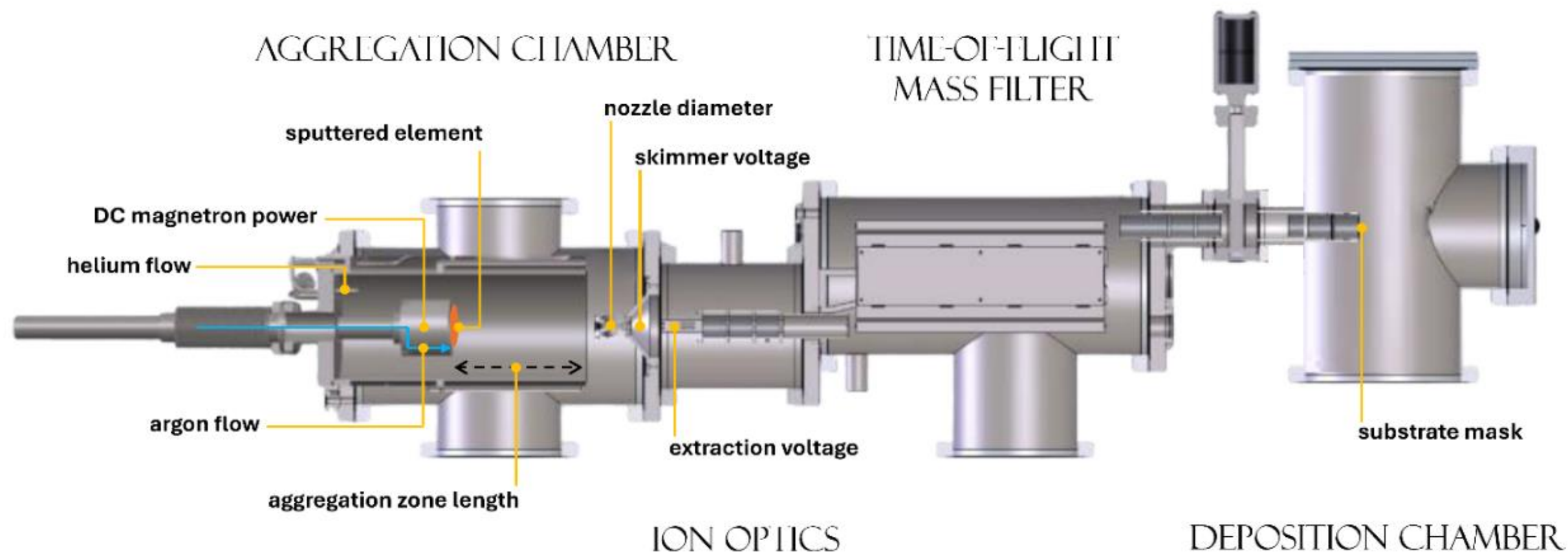

**Figure 1. Schematic representation of cluster beam deposition apparatus.**

The system employs a precision pressure control system, with the sputtering chamber connected to the external vacuum chamber via micro-nozzles with diameters ranging from 1 to 25 mm. During operation, the sputtering chamber maintains a relatively high-pressure environment of 0.01 – 0.1 Pa, while the external vacuum chamber is kept at a low pressure of $10^{-7}$ - $10^{-6}$ Pa. This significant pressure difference not only facilitates the condensation of sputtered atoms into clusters within the aggregation chamber but also drives the supersonic expansion of the gas carrying the particles as they pass through the exit nozzle. A beam diverter, positioned downstream, selectively removes excess carrier gas, allowing only the NP-rich central portion of the beam to enter the subsequent deposition chamber in a collision-free manner. Depending on experimental requirements, additional functional modules such as mass filtering or ion-optics processing can be flexibly integrated into the system. In the current study, however, where the focus was on the effect of deposition parameters on NP sizes, the time-of-flight apparatus was naturally not used as a mass filter but solely as a mass spectrometer. Specific deposition parameters are discussed in the Results and Discussion section. Post-deposition NP characterisation was performed on a Thermo Fisher Scientific Talos F200X transmission electron microscope operated at 200 kV.

### 2.2. Multivariate Linear Regression Analysis

Multiple linear regression is a statistical method used to model the relationship between a continuous response variable (in our case, NP sizes and size distributions) and multiple predictor variables (here, deposition parameters) [22-23]. The method estimates regression coefficients by minimising the sum of squared residuals, thereby quantifying the expected change in the response associated with a one-unit change in a predictor while holding all other predictors constant. This framework allows the separation of individual predictor effects in the presence of potentially correlated covariates. Continuous predictors (e.g., DC magnetron power) are often standardised to facilitate comparison of effect sizes and improve numerical stability, while categorical predictors (e.g., sputter-atom species) are incorporated using indicator variables relative to a reference level. Statistical inference is performed using t-tests on individual coefficients, and model adequacy is assessed through diagnostic evaluation of residuals, including checks for linearity, homoscedasticity, and approximate normality.

A common problem with identifying individual MS-IGC deposition parameter contributions to the produced output is the fact that often such parameters are not independent but rather intertwined. For example, it is generally established that the residence time of nascent NPs inside the condensation chamber is strongly affecting their size [24]. However, residence time depends on various input parameters all acting simultaneously, such as condensation zone length, inert-gas flows, pressure differential between the condensation and deposition chambers, etc. To this end, residual analysis in a multivariable context is used to examine structure in response variables that remains unexplained after accounting for measured predictors. By extracting residuals from fitted regression models and evaluating their correlation, it is possible to assess whether responses exhibit dependence beyond that attributable to shared covariates. A substantial residual correlation indicates the presence of latent or unmeasured factors influencing multiple responses simultaneously. This type of analysis complements both univariate regression and MANOVA by distinguishing predictor-driven associations from intrinsic coupling within the response system [25] .

Multivariate analysis of variance (MANOVA) extends the general linear model to situations involving multiple correlated response variables. Rather than analysing each response independently, MANOVA evaluates whether predictors produce systematic changes in the joint response space by comparing multivariate means across predictor values. This approach explicitly accounts for covariance among responses and is therefore more powerful and appropriate when responses are coupled. Test statistics such as Pillai's trace quantify the proportion of variance in the multivariate response explained by each predictor, after adjusting for all others. Pillai's trace is commonly preferred due to its robustness to violations of multivariate normality and its stability under moderate predictor collinearity. MANOVA provides an integrated assessment of predictor importance that cannot be obtained from separate univariate analyses alone.

The analysis was conducted on a dataset including repeated observations of a system characterised by multiple predictor variables and two continuous response variables, $y_1$ and $y_2$. The predictors included seven continuous variables ($x_1 - x_7$) and two originally discrete variables ($x_8$ and $x_9$). Variable $x_9$ (usage of mask at deposition chamber) was treated as a categorical factor throughout the analysis, while variable $x_8$ (sputtered atom species), initially encoded as a discrete factor (Ta or Cu), was subsequently replaced by a continuous analogue $x_{8\text{-}cont}$ (sputter threshold value per atomic species, indicating sputter yield; high threshold value corresponding to low yield) to assess the robustness of model results to predictor parameterisation. This dual encoding strategy allowed evaluation of whether conclusions were sensitive to the discretisation of experimental variables.

For the predictor variables, the following experimental ranges are assumed, reflecting the operational space of the system:

- DC Magnetron Power ($x_1$): 20 – 47 W
- Ar Flow ($x_2$): 15 – 45 sccm
- He Flow ($x_3$): 0 – 80 sccm
- Aggregation Zone Length ($x_4$): 100 – 200 mm
- Nozzle Diameter ($x_5$): 4 – 20 mm
- Skimmer Voltage ($x_6$): 60 – 170 V
- Extraction Voltage ($x_7$): 110 – 195 V
- Sputtered Species ($x_8$): Ta or Cu (replaced by $x_{8\text{-}cont}$: their corresponding sputter threshold values, 29.25 and 15.80 eV, respectively, from $Ar^+$ 1000 eV projectiles [26])
- Mask ($x_9$): On or Off

The response variables spanned wide dynamic ranges:

- NP Size ($y_1$): 450 – 6390 atoms/cluster
- NP Size FWHM ($y_2$): 615 – 6078 atoms/cluster

Continuous predictors ($x_1 – x_7$ and the continuous analogue of $x_8$) were standardised to zero mean and unit variance prior to model fitting. Standardisation was applied to facilitate direct comparison of regression coefficients, improve numerical conditioning, and ensure that multivariate test statistics were not dominated by differences in predictor scale. Categorical predictors were encoded as factors using treatment contrasts, with one level serving as the reference category and its effect absorbed into the intercept. The discrete parameter $x_8$ was replaced by a continuous variable, $x_{8\text{-}cont}$; the replacement was implemented by re-reading the original dataset and re-estimating all models using the updated predictor set. No other preprocessing steps were altered, ensuring that observed differences between models reflected only the change in predictor encoding rather than downstream analytical choices.

Separate multiple linear regression models were fitted for each response variable. The general model form was:

$$y_k = \beta_0 + \sum_{i=1}^{7} \beta_i x_i + \beta_8 x_8 + \beta_9 x_9 + \varepsilon_k \quad (1)$$

where $k \in \{1,2\}$, $x_9$ represents the categorical predictor, and $x_8$ represents either the discrete or continuous encoding depending on the model version. Regression assumptions were evaluated using residual-versus-fitted plots and quantile–quantile plots of standardised residuals. Multicollinearity among predictors was assessed using variance inflation factors; all continuous predictors exhibited acceptable values, indicating that collinearity was present but not severe enough to compromise coefficient stability. To assess whether the two responses exhibited dependence beyond that explained by the predictors, residuals from the fitted regression models were extracted and their Pearson correlation was computed. This analysis was used to identify latent coupling between $y_1$ and $y_2$ that was not attributable to shared predictor effects.

As $y_1$ and $y_2$ were found to be correlated, multivariate analysis of variance was employed to evaluate joint effects of predictors on the bivariate response. MANOVA models were fitted using the same predictor structure as the univariate regressions, thereby ensuring consistency across analytical frameworks. Pillai's trace was selected as the primary test statistic due to its robustness to deviations from multivariate normality and its favourable behaviour under moderate collinearity. Wilks' lambda was computed as a secondary check and yielded consistent inferential conclusions. To evaluate the effect of predictor encoding, MANOVA was performed both with $x_8$ treated as a categorical variable and with $x_8$ replaced by its continuous analogue. Comparisons focused on changes in multivariate effect size, predictor ranking, and statistical significance, providing a direct test of the robustness of multivariate conclusions to predictor parameterisation. To further characterise the dependence structure between $y_1$ and $y_2$, principal component analysis (PCA) was conducted on the standardised response matrix. PCA was used to identify dominant modes of variation and to determine whether system behaviour could be represented in a lower-dimensional response space. All analyses were performed using R (RStudio 2022.02.3+492) [27].

## 3. Results and Discussion

A typical MS-IGC source operates by accelerating ions within the process gas (typically Ar) and bombarding them onto a solid target (the cathode). When Ar ions collide with the target surface, they trigger the ejection of atoms, ions, and electrons, a **process known as sputtering.** To enhance sputtering efficiency, the device employs a permanent magnet (or array of magnets) to redirect ejected electrons back toward the target surface, creating a confined, helical magnetic field. To promote NP formation, a second inert gas (usually He) may be introduced into the aggregation chamber as a condensing medium and carrier gas. It should be emphasised that, to enable NP formation, the overall pressure in a MS-IGC aggregation chamber is typically higher than in common magnetron-sputtering sources for film deposition [4]. Residence time of the nascent clusters in the aggregation zone is determined by the zone's length (controlled by the exact position of the magnetron, which is mounted on a linear translation system), as well as the pressure differential between the aggregation and the deposition chambers (controlled by individual pumping systems, as well as by inert-gas flow rates and extraction efficiencies via the nozzle size and voltage). Such devices exhibit high efficiency due to the abundant dimer structures in the sputtered products, effectively overcoming the energy barrier constraints in the initial nucleation step of cluster formation [28]. Furthermore, approximately 50% of the generated NPs carry charge, enabling convenient integration of magnetron sputtering sources with mass selection and deposition techniques [29]. This method has been successfully applied to prepare various NP systems, typically yielding NPs ranging from <1 to >60 nm in diameter [30-34]. The size distribution can be controlled by adjusting parameters such as sputtering power, flow rates of sputtering and condensing gases, condensation chamber length, temperature, and end-aperture size (nozzle diameter) [20].

### 3.1. Nanoparticle Depositions

A total of 139 depositions of monometallic Cu or Ta NPs were performed, where all related deposition parameters were recorded and tabulated in Table S1. The depositions were monitored *operando* via the produced beam current, since all deposited NPs carry electrical charge. Neutral NPs, which may also be produced with MS-IGC, are not directed toward the deposition chamber, constituting what is often called the "white beam"; as such, they do not contribute to the measurements. All produced beam currents vs. NP sizes (expressed in number of atoms/cluster) are shown in Figure S1-S2 for Cu and Ta, respectively.

Typically, the size distribution of NPs fabricated by MS-IGC follow a log-normal distribution, which is, in principle, a normal distribution skewed by a relatively small number of outliers in the large-size regime. Such large-size NPs are formed due to the stochastic nature of the nucleation and growth mechanism, typically as a result of primary particle coalescence at a late stage of the process [10,35,36]. Provided the NPs are spherical and crystalline, packed in a similar fashion as their respective bulk materials, the conversion from atoms/cluster to diameter (in nm) is straightforward, following the general equation [37]:

$$d = \left(\frac{6NM}{\pi\,\rho\,N_A}\right)^{1/3} \tag{1}$$

where $d$ is the NP diameter, $N$ is the number of atoms/cluster, $M$ is the mass, $\rho$ is the density, and $N_A$ is Avogadro's number. The atomic packing, lattice constant, and atomic radius are implicitly included in equation (1) through the density. Substituting numerical values for Ta and Cu gives

$$d_{nm}^{Ta} \approx \left(\frac{N}{146}\right)^{1/3} \text{and } \ d_{nm}^{Cu} \approx \left(\frac{N}{444}\right)^{1/3} \qquad \text{(2a) and (2b)}$$

Thus, equal numbers of atoms correspond to different NP diameters for different materials. However, these conversions become less reliable at small NP sizes (< ~3 nm), which is within the range of the deposited NPs: therefore, the discussion on NPs size will be restricted to number of atoms/cluster. In a sense, this allows for direct comparison of the effect of different deposition parameters without having to deconvolute the results based on different atomic sizes or crystallographic systems (since, under ambient conditions, Ta is bcc whereas Cu is fcc), or having to account for post-deposition NP ripening occurring on the substrate (see Figure S3).

Exemplar size distributions of Ta and Cu NPs are shown in Figure 2, after gaussian smoothing, demonstrating log-normal dependency. Size distributions with less regular shapes also occurred, as shown in Figures S1-S2 of Supplementary Information, where lower, secondary peaks also formed at large NPs sizes. Therefore, to obtain consistent measurements, serving as representative of the produced NPs, fitting the experimental data with a log-normal function was avoided. Instead, the average NP sizes as the middle points of the full-widths at half-maxima (FWHM, indicated in the figure by the blue dash-dot lines) was assumed. The latter, being characteristic of the breadth of the NP size distributions, provided a measure of size homogeneity in each deposit, which was the second response variable investigated in our study.

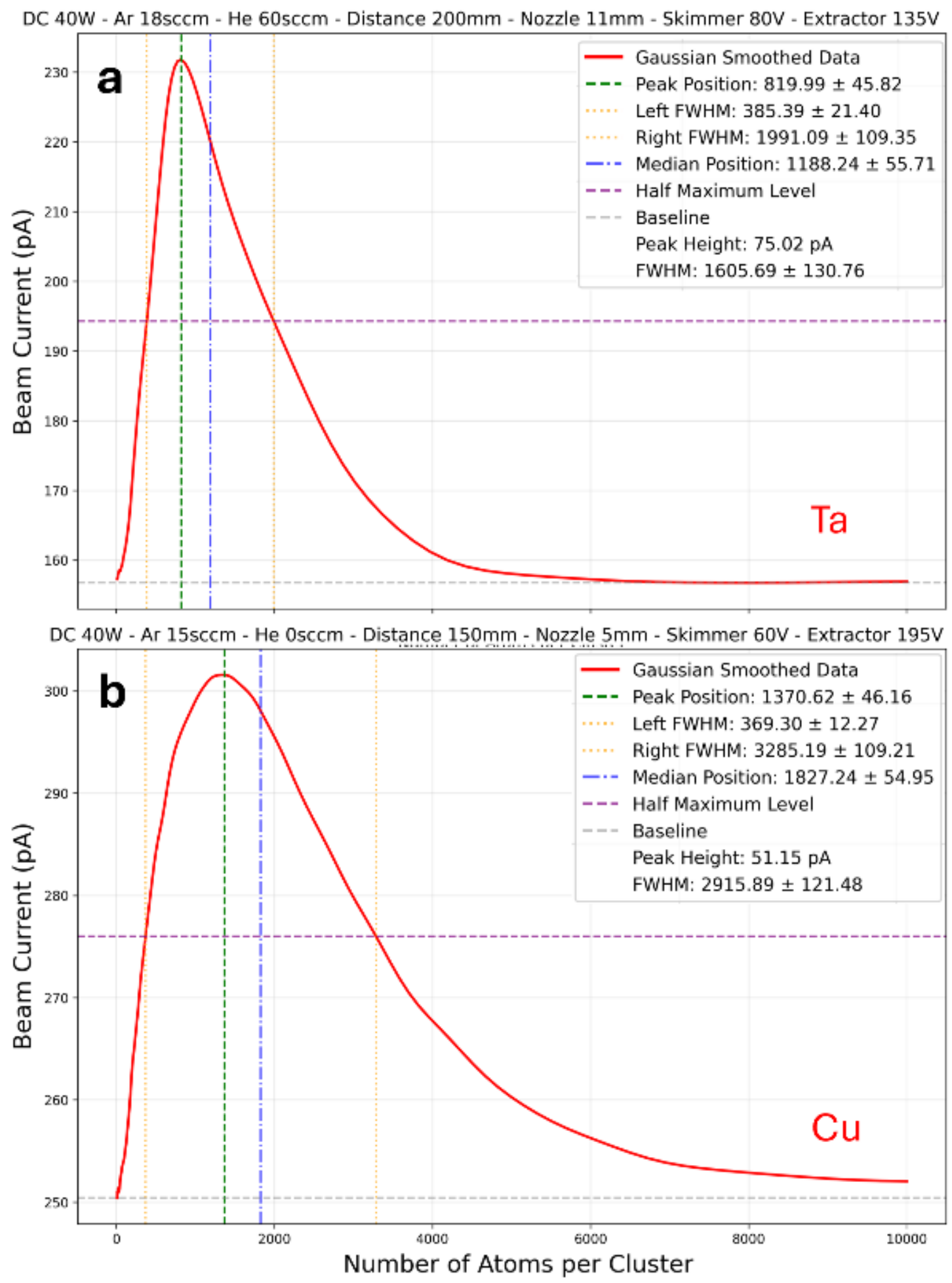

**Figure 2. Exemplar size distributions of Ta and Cu NPs (in atoms/cluster) deposited under prescribed deposition parameters.** Clearly the size distribution of Cu is wider than that of Ta, as indicated by their respective FWHM values. Peak positions corresponding the NPs of the most common sizes are indicated by the green dotted lines; median positions corresponding to the NPs average sizes are indicated by the blue dot-dashed lines.

## 3.2. Comparison of Effects of Different Deposition Parameters

### 3.2.1. Univariate Regression Results

The regression models for both NP average sizes and size distribution widths ($y_1$ and $y_2$, respectively) exhibited strong explanatory power, indicating that the predictor set captured a substantial fraction of the observed variability in both responses. For NP size, the model explained approximately 89% of the observed variance (adjusted $R^2 \approx 0.882$), while the model for FWHM explained approximately 85% (adjusted $R^2 \approx 0.844$). In both models, nozzle diameter ($x_5$) emerged as the strongest continuous predictor, exhibiting a large standardised effect with high statistical significance. Additional significant predictors included Ar flow, aggregation zone length, and extraction voltage ($x_2$, $x_4$, and $x_7$, respectively) while several predictors (magnetron power, He flow rate, and skimmer voltage-$x_1$, $x_3$, and $x_6$, respectively) did not contribute significantly once other variables were accounted for. The categorical predictor $x_9$ (mask on or off) showed a moderate effect in the model for NP size and a weaker effect for FWHM.

Replacing the discrete encoding of $x_8$ (namely, type of species) with its continuous analogue $x_{8\text{-}cont}$ (sputter threshold energy) did not substantially alter regression coefficients for any other predictors. Signs, magnitudes, and statistical significance of coefficients remained unchanged, indicating that the categorical version primarily acted as a coarse representation of an underlying continuous effect. The continuous analogue $x_{8\text{-}cont}$ did not exhibit statistically significant unique effects in either univariate model. Detailed results for each response are summarised below:

**Results for NP average size ($y_1$):** Model fit was strong ($R^2 = 0.889$, adjusted $R^2 = 0.882$). Statistically significant predictors included $x_2$ (positive), $x_4$ (positive), $x_5$ (strong negative), $x_7$ (negative), and $x_9$ (large categorical effect). Predictors $x_1$, $x_3$, $x_6$, and the continuous analogue $x_{8\text{-}cont}$ were not statistically significant.

**Results for FWHM ($y_2$):** Model fit was similarly strong ($R^2 = 0.854$, adjusted $R^2 = 0.844$). Significant predictors included $x_2$ (positive), $x_4$ (positive), $x_5$ (strong negative), $x_7$ (strong negative), and $x_9$, which produced the largest single shift in magnitude. Predictors $x_1$, $x_3$, $x_6$, and $x_{8\text{-}cont}$ were not statistically significant.

Figure 3 summarises these results by comparing standardised regression coefficients for $y_1$ and $y_2$ with the corresponding multivariate effects derived from MANOVA. Horizontal bars represent the relative magnitude and direction of predictor effects, scaled within each series to the interval [−1,1] to facilitate visual comparison, while numeric labels denote the corresponding unscaled regression coefficients or F-statistics and may therefore exceed the plotted axis range. The figure highlights the dominant influence of nozzle diameter ($x_5$) across both individual and joint responses, as well as the consistent contributions of the aggregation zone length and extraction voltage ($x_4$ and $x_7$, respectively) while also illustrating the comparatively weaker effects of the remaining predictors.

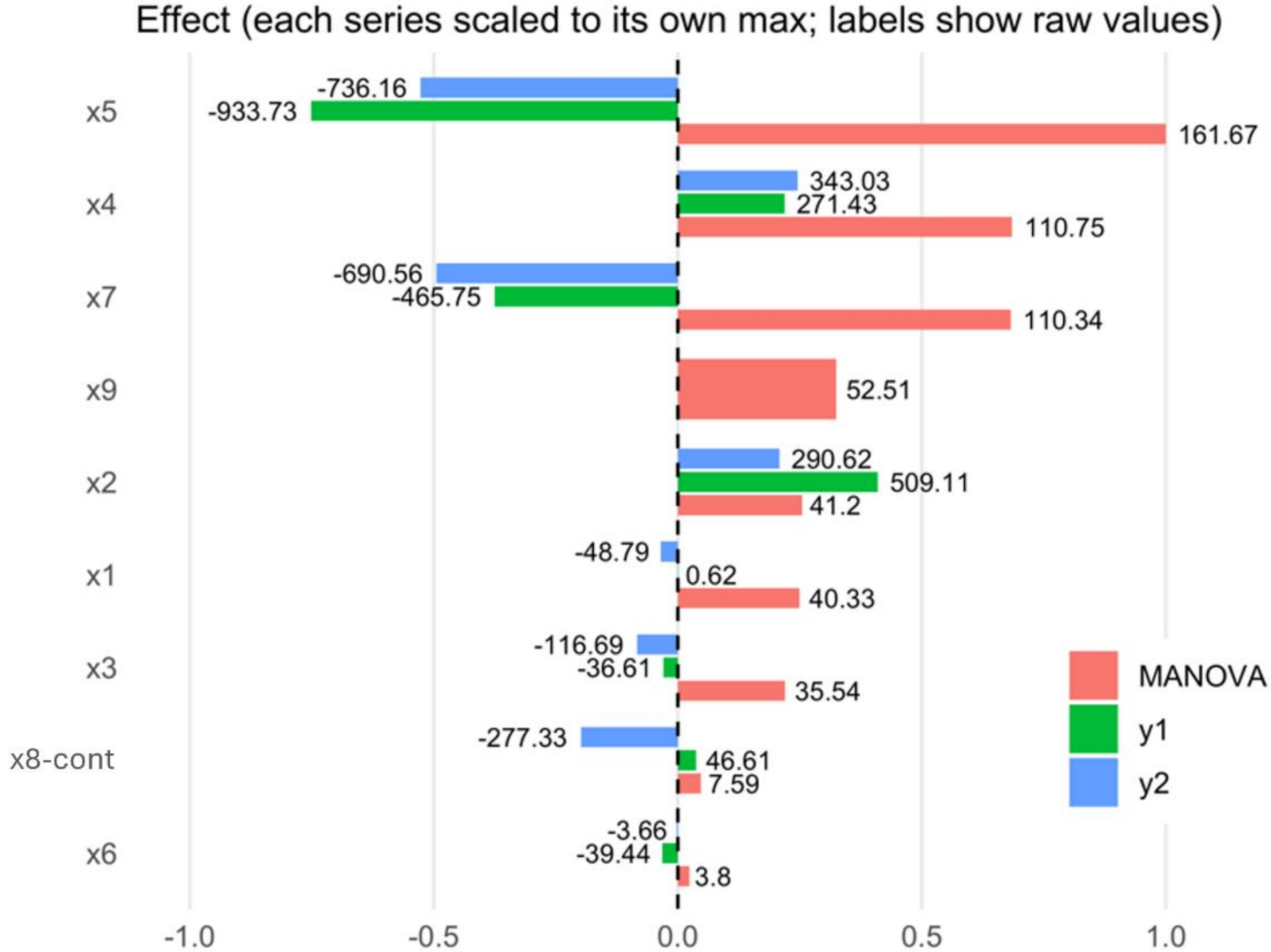

**Figure 3. Relative importance of predictors for the joint and individual responses.** Horizontal bars show standardised effects of each predictor on $y_1$ (blue) and $y_2$ (green) obtained from multiple linear regression, together with their joint multivariate effect from MANOVA (red; Pillai's trace expressed as the associated approximate F-statistic). For visual comparability, bar lengths are normalised within each series to the maximum absolute value in that series, such that all bars lie within the [−1,1] interval. Numeric labels indicate the corresponding raw regression coefficients or MANOVA F-statistics and are therefore not scaled. Positive and negative values denote the direction of the effect relative to the response. Reference levels of categorical predictors are omitted, as their effects are absorbed into the intercept. This representation highlights differences in both magnitude and direction of predictor effects across individual responses and their combined multivariate behaviour.

### 3.2.2. Residual Correlation Between Responses

Despite the high explanatory power of the regression models, a strong positive residual correlation was observed between $y_1$ and $y_2$ (NP size and size distribution) after accounting for all predictors ($r \approx 0.79$). This value indicates substantial shared latent structure influencing both responses that is not captured by the measured predictors. Importantly, the magnitude of this residual correlation remained stable across predictor encodings, suggesting that the observed coupling reflects intrinsic system behaviour rather than an artifact of variable parameterization. Figure 4 illustrates this residual coupling, showing a clear linear association between residuals of the two responses.

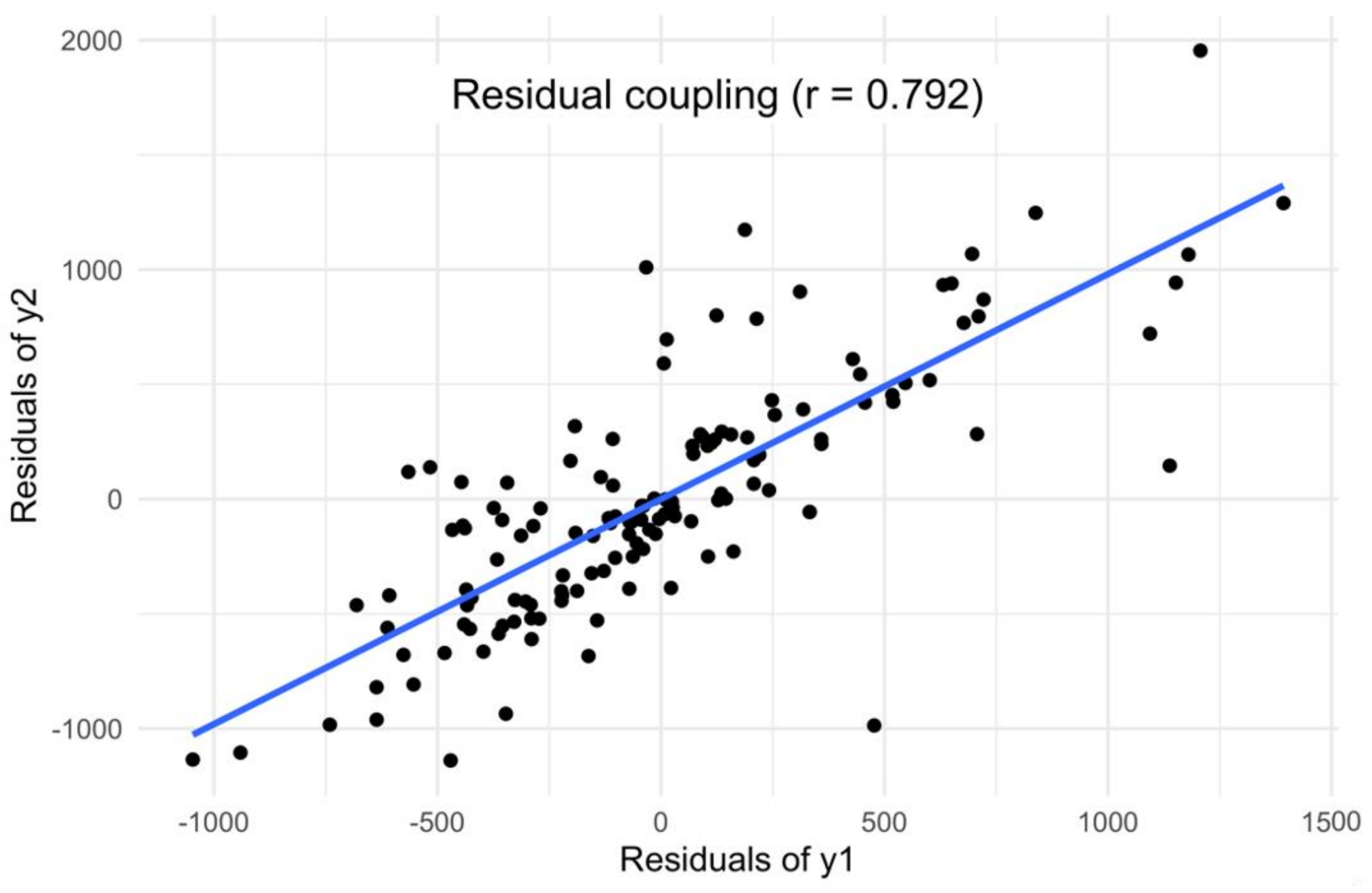

**Figure 4. Residual coupling between the two response variables after accounting for predictors.** Scatter plot of residuals from the multiple linear regression models for $y_1$ and $y_2$, fitted using the same set of predictors. Each point represents one observation; the solid line shows the least-squares fit to the residuals. A strong positive residual correlation (r=0.79) remains after controlling for all predictors, indicating substantial shared variation between the two responses that is not explained by the measured covariates. This residual dependence motivates the use of multivariate analysis (MANOVA) to assess joint predictor effects on the coupled response system.

### 3.2.3. Multivariate Effects on the Joint Response

MANOVA revealed that most predictors exerted statistically significant joint effects on the bivariate response ($y_1, y_2$). Across both predictor encodings, $x_5$ exhibited the largest Pillai trace, identifying nozzle diameter as the most influential continuous predictor for the joint behaviour of the two responses. Predictors $x_4$ and $x_7$ (aggregation zone length and extraction voltage) also demonstrated strong multivariate effects, while the effect of $x_6$ (skimmer voltage) was weak but statistically detectable. Although the continuous analogue $x_{8\text{-}cont}$ (sputter threshold energy) was not significant in the univariate regressions, it produced a small but statistically significant multivariate effect, indicating a modest yet consistent shift in the joint response space. This discrepancy between univariate and multivariate inference underscores the importance of analysing the coupled response system directly rather than relying solely on separate regressions. Replacing the categorical encoding with its continuous analogue did not alter the ranking or magnitude of multivariate effects, confirming the robustness of the joint response structure to predictor parameterisation.

Table 1 summarises the multivariate effects of all predictors, ordered by decreasing Pillai's trace, while Figure 5 emphasises the similar negative effect of $x_5$ (nozzle diameter), the most significant predictor, on both response variables.

**Table 1. Multivariate effects of predictors on the joint response ($y_1$, $y_2$) based on MANOVA.** The table reports Pillai's trace and corresponding p-values for each predictor from the multivariate analysis of variance evaluating joint effects on $y_1$ and $y_2$. Predictors are ordered by decreasing Pillai's trace, which provides a robust measure of multivariate effect size. Larger values indicate a greater proportion of variance explained in the bivariate response space. The categorical predictor $x_8$ produces a strong shift in the joint response, while $x_5$ emerges as the dominant continuous predictor. P-values reflect the significance of each predictor's unique contribution after accounting for all others in the model.

| Predictor | Deposition Parameter | Pillai | p-value | Interpretation |
|---|---|---|---|---|
| $x_5$ | Nozzle Diameter | 0.715 | < 2e-16 | Strongest multivariate effect |
| $x_4$ | Aggregation Zone Length | 0.632 | < 2e-16 | Very strong |
| $x_7$ | Extraction Voltage | 0.631 | < 2e-16 | Very strong |
| $x_9$ | Mask | 0.449 | < 2e-16 | Strong categorical shift |
| $x_2$ | Ar Flow | 0.390 | < 2e-14 | Moderate |
| $x_1$ | DC Magnetron Power | 0.385 | < 2e-14 | Moderate |
| $x_3$ | He Flow | 0.355 | < 2e-12 | Moderate |
| $x_{8\text{-}cont}$ | Sputter Threshold Energy | 0.105 | < 0.001 | Small but significant |
| $x_6$ | Skimmer Voltage | 0.056 | 0.025 | Weak |

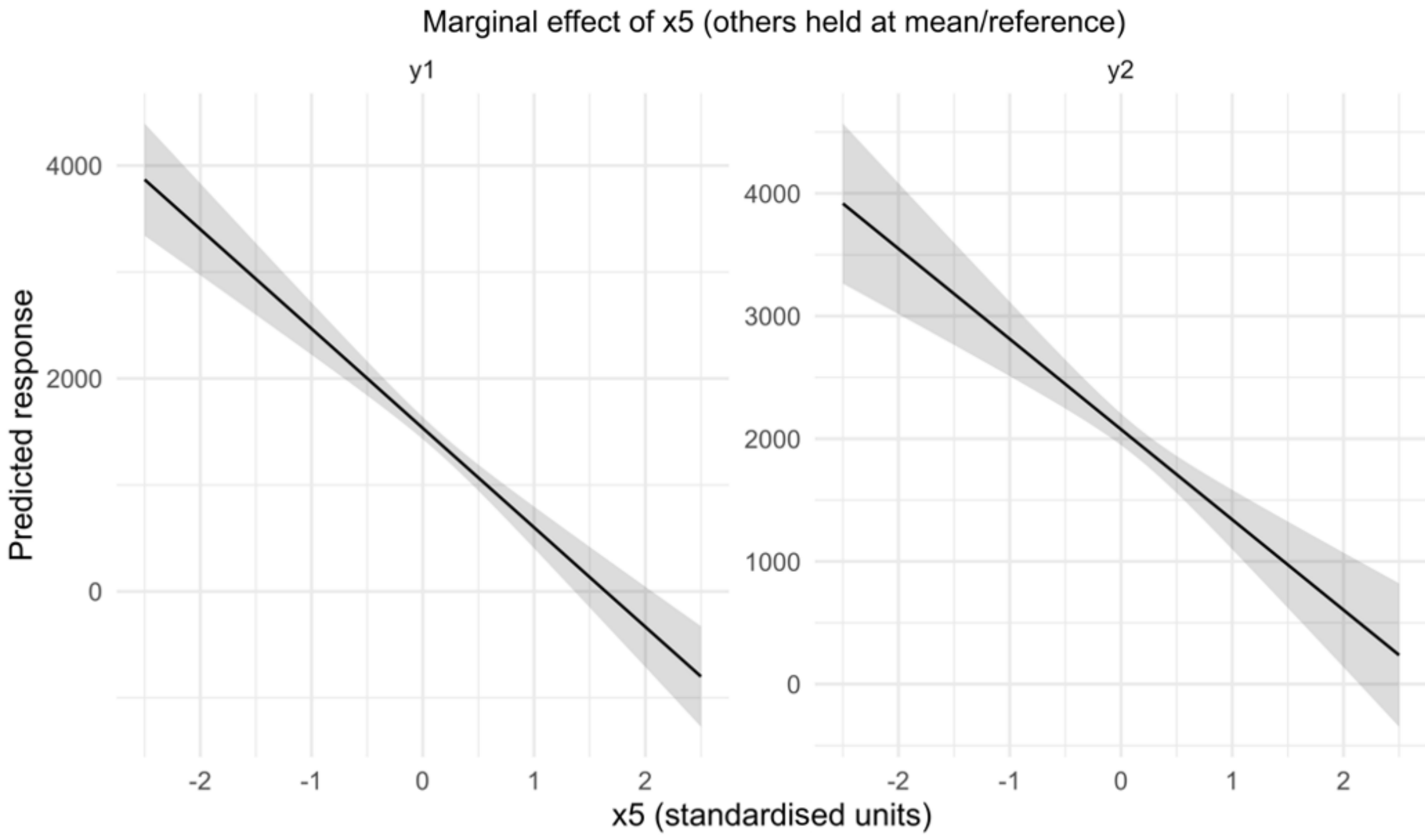

**Figure 5. Marginal effect of the dominant predictor $x_5$ (nozzle diameter) on each response variable.** Predicted values of $y_1$ (left panel) and $y_2$ (right panel) as a function of the standardised predictor $x_5$, obtained from the fitted multiple linear regression models. Solid lines indicate model predictions, while shaded bands denote 95% confidence intervals. All other continuous predictors were held at their mean values (zero after standardisation), and categorical predictors were fixed at their reference levels. The consistently negative slope in both panels indicates that increases in $x_5$ are associated with decreases in both responses, confirming $x_5$ as the dominant continuous predictor influencing the coupled response system.

#### 3.2.4. Principal Component Structure of the Responses

Principal component analysis was performed on the standardised response matrix to characterise the intrinsic structure of the bivariate response space. The first principal component (PC1) accounted for 97.7% of the total variance, while the second component (PC2) explained only 2.3%, indicating an overwhelmingly dominant single mode of variation. The loading structure revealed that PC1 consisted of equal positive contributions from both responses (loading ≈ +0.707 for both $y_1$ and $y_2$), corresponding to a common system-wide response mode in which both variables increase or decrease together. In contrast, PC2 represented an orthogonal contrast between the responses (loading ≈ −0.707 for $y_1$ and +0.707 for $y_2$), capturing minor differential behaviour between them. The near-complete dominance of PC1 demonstrates that the two responses lie almost entirely along a single latent dimension. This finding is fully consistent with the strong residual correlation ($r \approx 0.79$) observed after regression adjustment and confirms that $y_1$ and $y_2$ form a tightly coupled response system governed primarily by a shared underlying process. The minimal variance explained by PC2 indicates that response-specific deviations from this common mode are comparatively small.

The concordance between PCA, residual coupling analysis, and MANOVA supports the interpretation that the observed low-dimensional structure reflects intrinsic system behaviour rather than model specification artifacts.

#### 3.2.5. Physical Meaning of Multiple Linear Regression Analysis

It has been well documented in the literature that a combination of gas composition (of both sputter and inert-gas atoms), pressure, and NP residence time in the aggregation chamber are key regarding the output NP sizes [38]. However, none of these parameters can be directly controlled by the experimentalist; instead, they are the result of a complex interplay between controllable deposition parameters that govern the structure of the final output sample. For example, if one wishes to increase the residence time, one may decrease the exit orifice diameter. However, this would also certainly affect the pressure inside the chamber. A simple empirical relationship exists between the aggregation chamber pressure and the residence time of the carrier gas that pushes the NPs toward the deposition chamber [24]:

$$t = \frac{\pi r^2 L P}{Q} \tag{3}$$

where $r$ and $L$ are the chamber's radius and length, respectively (assuming it is cylindrical, which is usually the case, and measuring $L$ as the distance between the sputter gun and the orifice), $P$ is the pressure inside the chamber, and $Q$ is the carrier gas flow rate. Nevertheless, this equation does not account for additional complications such as affected fluid dynamics within the chamber, or, potentially, size-dependent changes in the mean free path of nascent NPs between collisions with each other or inert-gas atoms. As a result, although it is reasonable to assume that larger NPs are thus produced due to prolonged nucleation and growth and extensive coalescence, it is practically impossible to deconvolute whether this is a result of residence time in itself, or of pressure increase, etc. Therefore, this study is focused on the controllable deposition parameters (aforementioned predictor variables) alone.

It is clear from our statistical analysis that the dominant deposition parameters are, indeed, those most closely related to residence time: nozzle diameter, aggregation zone length, and extraction voltage, in agreement with previous reports [39,40]. Parameters affecting the density and composition of the sputter plasma (magnetron power, Ar and/or He gas flow rates) appear to have a rather moderate effect on NP size. The combination of these two findings implies that a substantial part of NP growth occurs at a distance from the sputter target, supporting the two-growth zones hypothesis [24,41]. As expected, conditions that favour larger NP sizes also allow for broader size distributions, thus contributing to the correlation between the two response variables. The presence of a mask near the substrate also has a statistically significant effect, indicating that its masking efficiency may be size-dependent.

It should be noted that nanoparticle nucleation and growth in gas-phase aggregation sources are inherently nonlinear processes involving complex interactions between plasma conditions, collision dynamics, and transport phenomena. The linear regression framework adopted here is therefore best interpreted as a local approximation within the experimentally explored parameter space rather than a universal description of nanoparticle formation. The strong explanatory power of the models nevertheless indicates that first-order linear effects dominate the observed variability over the investigated operating window.

In closing, it should be noted that although this study is relatively extensive, its conclusions are valid for the parameter ranges studied. Should another study go significantly beyond these ranges, there is always a possibility that some of the discovered dependencies may change. For example, magnetron power is positively correlated with NP size, through the supersaturation of sputter atoms and its resultant effect on nucleation and growth kinetics [20]. However, an increase beyond certain values even within operating conditions (that is, before plasma arching or target destruction) might result in decrease in this supersaturation due to various factors such as excessive kinetic energy of the sputtered atoms, redeposition of nascent NPs on the target, etc.

## 4. Conclusions

This work provides a quantitative examination of how controllable process parameters in MS-IGC influence NP size and size distribution. By combining a large experimental dataset of 139 depositions with multiple linear regression and multivariate statistical analysis, the study isolates the relative contribution of individual deposition parameters while accounting for their simultaneous effects. The results demonstrate that parameters governing the effective residence time of nascent clusters—particularly nozzle diameter, aggregation zone length, and extraction voltage—dominate the control of both NP size and size distribution width. In contrast, parameters primarily associated with plasma generation and gas composition (magnetron power and inert-gas flows) exhibit more moderate influence once other variables are considered. These findings support the widely proposed but rarely quantified view that residence-time-related mechanisms play the central role in determining particle growth dynamics in MS-IGC systems.

Beyond identifying dominant predictors, the statistical framework reveals that NP size and size distribution are strongly coupled responses that evolve along a common underlying growth mode. Residual correlation analysis, MANOVA, and principal component analysis consistently show that most variability in the system can be described along a single latent dimension, indicating that conditions favouring larger particles simultaneously broaden the size distribution. Importantly, the ranking and magnitude of parameter effects remain robust under different predictor encodings, demonstrating that the observed relationships reflect intrinsic system behaviour rather than model artefacts.

Overall, the present analysis illustrates how multivariate statistical tools can complement physical intuition to disentangle the complex interplay of MS-IGC deposition parameters, offering a practical framework for guiding experimental optimisation and for improving predictive control of gas-phase nanoparticle synthesis. Such tools could be most effective in more comprehensive future studies incorporating wider values spectra, as well as data from different laboratories (including those capable of depositing and probing neutral nanoparticle populations), various apparatus geometries (multiple sources, gas inlet positions, etc.), and including more deposition and response parameters (such as plasma balance, aggregation chamber shape, inclusion of other gasses, or NP shapes, cover densities, etc., respectively).

Finally, the present study deliberately prioritised interpretability and physical insight over predictive complexity. While nonlinear models may achieve marginal improvements in predictive accuracy, the linear framework allows direct identification and ranking of process parameters, facilitating physical interpretation of nanoparticle growth mechanisms. Future studies employing nonlinear regression, interaction terms, or machine-learning approaches may provide additional insight into higher-order dependencies and parameter coupling effects.

**CRediT authorship contribution statement**

**Y. Wang:** Experiment, Formal analysis, Writing – original draft. **Abisegapriyan K.S.:** Experiment, Supervision, Formal analysis. **E. Toulkeridou:** Formal analysis, Writing-original draft. **Y. Ein-Eli:** Supervision, Writing – review & editing, **P. Grammatikopoulos:** Writing – original draft, review & editing, Formal analysis, Supervision, Resources, Project administration, Methodology, Conceptualisation.

**Declaration of competing interest**

The authors declare no conflict of interest.

**Data availability**

All data are available upon request.

**Acknowledgements**

PG was supported by a Beatriz Galindo Senior Fellowship (BG23/00144). Work performed at GTIIT was supported by funding from the Guangdong Technion Israel Institute of Technology and Guangdong Provincial Key Laboratory of Materials and Technologies for Energy Conversion, MATEC (No. MATEC2022KF001). YEE work was supported by Technion's GTEP and INIES Institutes. PG is grateful to Prof S Pratsinis for his valuable advice and support during the setup of the experimental NanoGRaM Lab at GTIIT and his overall mentorship.

# SUPPLEMENTARY INFORMATION

**Table S1. Input (deposition) parameters and output response (nanoparticle size) variables.**

| | Deposition parameters (input predictor variables) | | | | | | | | | Nanoparticle sizes and distribution (output response variables) | | | | | |
|---|---|---|---|---|---|---|---|---|---|---|---|---|---|---|---|
| **Number** | **DC Magnetron Power (W)** | **Ar Flow (sccm)** | **He Flow (sccm)** | **Aggregation Zone Length (mm)** | **Nozzle Diameter (mm)** | **Skimmer Voltage (V)** | **Extraction Voltage (V)** | **Sputtered Species** | **Mask (on or off)** | **Peak Position (atoms/cluster)** | **Peak Position error (atoms/cluster)** | **FWHM (atoms/cluster)** | **FWHM error (atoms/cluster)** | **NP size (atoms/cluster)** | **NP size error (atoms/cluster)** |
| | $x_1$ | $x_2$ | $x_3$ | $x_4$ | $x_5$ | $x_6$ | $x_7$ | $x_8$ | $x_9$ | | | $y_2$ | | $y_1$ | |
| 1 | 20 | 18 | 60 | 200 | 11 | 80 | 135 | Ta | off | 531 | 30 | 1380 | 108 | 982 | 47 |
| 2 | 30 | 20 | 60 | 190 | 11 | 90 | 180 | Ta | off | 735 | 41 | 1198 | 102 | 926 | 43 |
| 3 | 30 | 20 | 60 | 190 | 11 | 90 | 180 | Ta | off | 625 | 35 | 1155 | 88 | 813 | 38 |
| 4 | 30 | 20 | 60 | 200 | 11 | 80 | 135 | Ta | off | 1267 | 71 | 2075 | 175 | 1611 | 73 |
| 5 | 30 | 20 | 60 | 200 | 11 | 80 | 180 | Ta | off | 965 | 54 | 1228 | 121 | 1136 | 49 |
| 6 | 32 | 28 | 60 | 180 | 15 | 90 | 180 | Ta | off | 377 | 44 | 1045 | 143 | 637 | 65 |
| 7 | 32 | 30 | 45 | 180 | 15 | 90 | 180 | Ta | off | 469 | 54 | 1087 | 140 | 636 | 65 |
| 8 | 32 | 30 | 50 | 180 | 15 | 90 | 180 | Ta | off | 583 | 67 | 1107 | 140 | 650 | 65 |
| 9 | 32 | 30 | 55 | 180 | 15 | 90 | 180 | Ta | off | 377 | 44 | 1043 | 142 | 625 | 65 |
| 10 | 32 | 30 | 60 | 180 | 14 | 90 | 180 | Ta | off | 377 | 44 | 1042 | 139 | 604 | 65 |
| 11 | 32 | 30 | 60 | 180 | 15 | 90 | 180 | Ta | off | 377 | 44 | 986 | 128 | 600 | 58 |
| 12 | 32 | 30 | 60 | 180 | 15 | 90 | 180 | Ta | off | 523 | 60 | 1036 | 144 | 652 | 65 |
| 13 | 32 | 30 | 65 | 180 | 15 | 90 | 180 | Ta | off | 338 | 39 | 922 | 127 | 559 | 58 |
| 14 | 32 | 30 | 70 | 180 | 15 | 90 | 180 | Ta | off | 377 | 44 | 978 | 125 | 565 | 58 |
| 15 | 32 | 30 | 75 | 180 | 15 | 90 | 180 | Ta | off | 338 | 39 | 937 | 127 | 561 | 58 |
| 16 | 32 | 30 | 80 | 180 | 15 | 90 | 180 | Ta | off | 338 | 39 | 869 | 115 | 529 | 52 |
| 17 | 32 | 32 | 60 | 180 | 15 | 90 | 180 | Ta | off | 377 | 44 | 914 | 128 | 564 | 58 |
| 18 | 32 | 34 | 60 | 180 | 15 | 90 | 180 | Ta | off | 377 | 44 | 838 | 121 | 562 | 53 |
| 19 | 32 | 36 | 60 | 180 | 15 | 90 | 180 | Ta | off | 377 | 44 | 929 | 129 | 585 | 59 |
| 20 | 35 | 20 | 30 | 200 | 11 | 90 | 130 | Ta | on | 1941 | 224 | 3356 | 575 | 2570 | 245 |
| 21 | 33 | 20 | 30 | 200 | 11 | 90 | 130 | Ta | on | 2165 | 250 | 2846 | 536 | 2389 | 222 |
| 22 | 33 | 20 | 30 | 200 | 11 | 90 | 130 | Ta | on | 2165 | 250 | 3301 | 598 | 2697 | 248 |
| 23 | 33 | 22 | 40 | 175 | 11 | 90 | 130 | Ta | on | 2165 | 250 | 3680 | 557 | 2506 | 244 |
| 24 | 33 | 24 | 20 | 200 | 11 | 90 | 130 | Ta | on | 3005 | 347 | 4697 | 830 | 3748 | 344 |
| 25 | 33 | 24 | 40 | 175 | 11 | 90 | 130 | Ta | on | 1941 | 224 | 3142 | 516 | 2375 | 220 |
| 26 | 33 | 24 | 40 | 175 | 11 | 90 | 130 | Ta | on | 1941 | 224 | 4994 | 667 | 3139 | 302 |
| 27 | 33 | 22 | 40 | 185 | 11 | 90 | 130 | Ta | on | 2165 | 250 | 3738 | 575 | 2759 | 245 |
| 28 | 33 | 24 | 40 | 200 | 11 | 90 | 130 | Ta | on | 2415 | 279 | 3575 | 598 | 2839 | 248 |
| 29 | 33 | 24 | 50 | 200 | 11 | 90 | 130 | Ta | on | 2165 | 250 | 3909 | 642 | 2862 | 274 |
| 30 | 33 | 28 | 30 | 175 | 12 | 90 | 130 | Ta | on | 2165 | 250 | 4480 | 693 | 3085 | 303 |
| 31 | 37 | 30 | 30 | 175 | 11 | 90 | 130 | Ta | on | 3352 | 387 | 4385 | 849 | 3857 | 346 |
| 32 | 38 | 30 | 30 | 180 | 15 | 90 | 150 | Ta | off | 810 | 94 | 1532 | 232 | 1046 | 102 |
| 33 | 44 | 30 | 50 | 180 | 15 | 170 | 180 | Ta | off | 726 | 84 | 1298 | 175 | 768 | 81 |
| 34 | 38 | 30 | 40 | 180 | 15 | 90 | 150 | Ta | off | 903 | 104 | 1482 | 203 | 958 | 91 |
| 35 | 40 | 18 | 60 | 200 | 11 | 80 | 135 | Ta | off | 820 | 46 | 1606 | 131 | 1188 | 56 |
| 36 | 40 | 23 | 40 | 185 | 11 | 90 | 120 | Ta | off | 2694 | 311 | 4618 | 683 | 3082 | 303 |
| 37 | 40 | 23 | 55 | 185 | 11 | 90 | 120 | Ta | off | 2415 | 279 | 3645 | 557 | 2543 | 244 |
| 38 | 40 | 23 | 60 | 185 | 11 | 90 | 120 | Ta | off | 2415 | 279 | 3794 | 566 | 2676 | 244 |
| 39 | 43 | 26 | 30 | 180 | 9 | 90 | 140 | Ta | on | 3005 | 347 | 4019 | 761 | 3408 | 310 |
| 40 | 40 | 26 | 40 | 180 | 11 | 90 | 130 | Ta | on | 1941 | 224 | 3828 | 642 | 2821 | 274 |
| 41 | 40 | 26 | 40 | 190 | 11 | 90 | 130 | Ta | on | 2165 | 250 | 3579 | 598 | 2840 | 248 |
| 42 | 40 | 26 | 40 | 200 | 11 | 90 | 130 | Ta | on | 2694 | 311 | 4088 | 729 | 3268 | 307 |
| 43 | 40 | 26 | 50 | 200 | 11 | 90 | 130 | Ta | on | 2415 | 279 | 3675 | 654 | 2928 | 275 |
| 44 | 40 | 27 | 30 | 200 | 11 | 90 | 130 | Ta | on | 3005 | 347 | 4206 | 761 | 3503 | 310 |
| 45 | 40 | 28 | 30 | 200 | 11 | 90 | 130 | Ta | on | 3005 | 347 | 4257 | 761 | 3573 | 310 |
| 46 | 40 | 29 | 30 | 200 | 11 | 90 | 130 | Ta | on | 3352 | 387 | 3986 | 780 | 3675 | 313 |
| 47 | 40 | 30 | 30 | 180 | 15 | 90 | 130 | Ta | off | 651 | 75 | 1576 | 224 | 1006 | 101 |
| 48 | 38 | 30 | 35 | 185 | 11 | 90 | 135 | Ta | off | 1008 | 116 | 1582 | 273 | 1219 | 115 |
| 49 | 40 | 30 | 35 | 185 | 15 | 90 | 135 | Ta | off | 810 | 94 | 1410 | 215 | 1021 | 92 |
| 50 | 40 | 30 | 35 | 185 | 15 | 90 | 150 | Ta | off | 523 | 60 | 997 | 152 | 696 | 66 |
| 51 | 40 | 30 | 35 | 185 | 15 | 90 | 180 | Ta | off | 377 | 44 | 810 | 119 | 538 | 53 |
| 52 | 40 | 30 | 40 | 185 | 15 | 90 | 180 | Ta | off | 338 | 39 | 831 | 119 | 542 | 53 |
| 53 | 40 | 30 | 50 | 180 | 15 | 90 | 150 | Ta | off | 469 | 54 | 900 | 139 | 655 | 59 |
| 54 | 40 | 30 | 50 | 180 | 15 | 90 | 180 | Ta | off | 338 | 39 | 651 | 100 | 471 | 43 |
| 55 | 40 | 30 | 50 | 185 | 15 | 90 | 180 | Ta | off | 377 | 44 | 703 | 110 | 493 | 47 |
| 56 | 40 | 30 | 60 | 180 | 15 | 90 | 180 | Ta | off | 377 | 44 | 684 | 108 | 481 | 47 |
| 57 | 40 | 30 | 30 | 175 | 11 | 90 | 130 | Ta | on | 3352 | 387 | 4287 | 849 | 3706 | 346 |
| 58 | 40 | 30 | 30 | 180 | 15 | 90 | 130 | Ta | on | 1941 | 224 | 3483 | 486 | 2230 | 218 |
| 59 | 40 | 30 | 30 | 185 | 11 | 90 | 130 | Ta | on | 3352 | 387 | 4660 | 849 | 4052 | 346 |
| 60 | 40 | 30 | 30 | 190 | 15 | 90 | 130 | Ta | on | 1560 | 180 | 2973 | 448 | 2059 | 196 |
| 61 | 40 | 30 | 30 | 190 | 11 | 90 | 130 | Ta | on | 3740 | 432 | 4562 | 849 | 4002 | 346 |
| 62 | 40 | 30 | 30 | 200 | 11 | 90 | 130 | Ta | on | 3352 | 387 | 4425 | 849 | 3886 | 346 |
| 63 | 40 | 30 | 40 | 160 | 12 | 110 | 120 | Ta | off | 726 | 84 | 1410 | 215 | 1021 | 92 |
| 64 | 40 | 30 | 40 | 180 | 12 | 110 | 120 | Ta | off | 726 | 84 | 2297 | 314 | 1474 | 141 |
| 65 | 40 | 30 | 40 | 180 | 15 | 90 | 135 | Ta | off | 903 | 104 | 1708 | 263 | 1210 | 114 |
| 66 | 40 | 30 | 40 | 170 | 15 | 120 | 130 | Ta | off | 383 | 45 | 1020 | 132 | 597 | 61 |
| 67 | 40 | 30 | 40 | 170 | 13 | 90 | 130 | Ta | off | 931 | 109 | 3004 | 417 | 1918 | 187 |
| 68 | 40 | 30 | 40 | 170 | 15 | 90 | 130 | Ta | off | 176 | 21 | 851 | 103 | 476 | 49 |
| 69 | 40 | 30 | 40 | 170 | 9 | 90 | 130 | Ta | off | 1623 | 191 | 2165 | 442 | 2040 | 175 |
| 70 | 40 | 30 | 50 | 160 | 15 | 120 | 130 | Ta | off | 534 | 63 | 983 | 139 | 638 | 62 |
| 71 | 40 | 30 | 50 | 170 | 15 | 120 | 130 | Ta | off | 343 | 40 | 809 | 109 | 496 | 49 |
| 72 | 43 | 32 | 30 | 180 | 9 | 90 | 120 | Ta | on | 4171 | 482 | 5807 | 1056 | 5016 | 431 |
| 73 | 40 | 28 | 40 | 175 | 11 | 90 | 120 | Ta | on | 2165 | 250 | 4886 | 785 | 3473 | 339 |
| 74 | 42 | 28 | 40 | 185 | 11 | 90 | 120 | Ta | on | 2415 | 279 | 4556 | 729 | 3449 | 307 |
| 75 | 42 | 28 | 40 | 185 | 11 | 90 | 120 | Ta | off | 2415 | 279 | 3831 | 654 | 2947 | 275 |
| 76 | 43 | 32 | 50 | 180 | 9 | 90 | 120 | Ta | on | 3740 | 432 | 4330 | 830 | 3721 | 344 |
| 77 | 40 | 40 | 40 | 170 | 9 | 90 | 130 | Ta | off | 2830 | 333 | 3155 | 710 | 3337 | 276 |
| 78 | 40 | 45 | 40 | 170 | 9 | 90 | 130 | Ta | on | 4933 | 580 | 6078 | 1382 | 6390 | 538 |
| 79 | 40 | 45 | 40 | 170 | 9 | 90 | 130 | Ta | off | 3949 | 464 | 4021 | 1022 | 4703 | 391 |
| 80 | 40 | 30 | 50 | 180 | 15 | 90 | 135 | Ta | off | 651 | 75 | 1635 | 224 | 1026 | 101 |
| 81 | 40 | 30 | 60 | 180 | 15 | 90 | 135 | Ta | off | 651 | 75 | 1563 | 229 | 1047 | 101 |
| 82 | 43 | 26 | 40 | 180 | 9 | 90 | 140 | Ta | on | 3005 | 347 | 3650 | 667 | 3071 | 276 |
| 83 | 41 | 30 | 30 | 190 | 12 | 90 | 110 | Ta | off | 3005 | 347 | 4232 | 761 | 3651 | 310 |
| 84 | 43 | 30 | 30 | 190 | 12 | 90 | 110 | Ta | off | 1740 | 201 | 4052 | 621 | 2782 | 272 |
| 85 | 43 | 30 | 30 | 190 | 12 | 90 | 130 | Ta | off | 1398 | 162 | 2852 | 448 | 1957 | 196 |
| 86 | 43 | 30 | 30 | 190 | 12 | 90 | 150 | Ta | off | 1254 | 145 | 2184 | 322 | 1473 | 141 |
| 87 | 42 | 30 | 40 | 180 | 15 | 90 | 150 | Ta | off | 523 | 60 | 1016 | 152 | 710 | 66 |
| 88 | 42 | 30 | 40 | 180 | 20 | 90 | 150 | Ta | off | 651 | 75 | 1238 | 190 | 880 | 82 |
| 89 | 43 | 30 | 50 | 180 | 15 | 110 | 180 | Ta | off | 583 | 67 | 902 | 155 | 691 | 66 |
| 90 | 43 | 30 | 50 | 180 | 15 | 130 | 180 | Ta | off | 469 | 54 | 729 | 127 | 562 | 53 |
| 91 | 43 | 30 | 50 | 180 | 15 | 150 | 180 | Ta | off | 469 | 54 | 768 | 127 | 590 | 53 |
| 92 | 43 | 30 | 50 | 180 | 15 | 90 | 180 | Ta | off | 377 | 44 | 615 | 100 | 450 | 43 |
| 93 | 43 | 28 | 30 | 175 | 9 | 90 | 120 | Ta | on | 3740 | 432 | 5149 | 947 | 4411 | 386 |
| 94 | 43 | 28 | 40 | 175 | 11 | 90 | 120 | Ta | on | 3352 | 387 | 4593 | 830 | 3709 | 344 |
| 95 | 43 | 28 | 40 | 175 | 11 | 90 | 120 | Ta | on | 2694 | 311 | 4237 | 631 | 3014 | 273 |
| 96 | 43 | 28 | 40 | 175 | 12 | 90 | 120 | Ta | on | 3005 | 347 | 4771 | 704 | 3342 | 304 |
| 97 | 43 | 28 | 40 | 175 | 9 | 90 | 120 | Ta | on | 3352 | 387 | 4283 | 761 | 3581 | 310 |
| 98 | 44 | 30 | 30 | 185 | 12 | 90 | 150 | Ta | off | 1124 | 130 | 2018 | 285 | 1332 | 126 |
| 99 | 44 | 30 | 30 | 190 | 12 | 90 | 150 | Ta | off | 1008 | 116 | 2098 | 318 | 1404 | 141 |
| 100 | 44 | 30 | 40 | 180 | 15 | 90 | 150 | Ta | off | 583 | 67 | 1011 | 150 | 699 | 66 |
| 101 | 45 | 30 | 50 | 160 | 15 | 120 | 130 | Ta | off | 833 | 98 | 1346 | 200 | 934 | 87 |
| 102 | 45 | 30 | 50 | 160 | 18 | 120 | 130 | Ta | off | 1163 | 137 | 2316 | 395 | 1699 | 169 |
| 103 | 45 | 28 | 30 | 175 | 9 | 90 | 120 | Ta | on | 3740 | 432 | 4358 | 893 | 4128 | 353 |
| 104 | 45 | 32 | 30 | 200 | 9 | 90 | 120 | Ta | on | 4171 | 482 | 5362 | 1056 | 4753 | 431 |
| 105 | 45 | 40 | 30 | 180 | 14 | 90 | 130 | Ta | on | 4209 | 476 | 4206 | 1008 | 4836 | 390 |
| 106 | 45 | 45 | 40 | 170 | 9 | 90 | 115 | Ta | on | 5512 | 648 | 4450 | 1320 | 5950 | 498 |
| 107 | 47 | 30 | 50 | 160 | 15 | 120 | 130 | Ta | off | 746 | 88 | 1183 | 179 | 814 | 77 |
| 108 | 20 | 20 | 0 | 100 | 5 | 60 | 195 | Cu | Off | 640 | 22 | 1579 | 68 | 1033 | 30 |
| 109 | 20 | 20 | 0 | 150 | 4 | 90 | 190 | Cu | Off | 1240 | 68 | 3200 | 181 | 1650 | 89 |
| 110 | 20 | 20 | 0 | 150 | 5 | 60 | 195 | Cu | Off | 984 | 33 | 2948 | 121 | 1842 | 55 |
| 111 | 30 | 15 | 0 | 100 | 5 | 60 | 195 | Cu | Off | 707 | 24 | 2145 | 85 | 1266 | 39 |
| 112 | 30 | 15 | 0 | 125 | 5 | 60 | 195 | Cu | Off | 1124 | 38 | 2729 | 107 | 1617 | 50 |
| 113 | 30 | 15 | 0 | 150 | 5 | 60 | 195 | Cu | Off | 1464 | 49 | 3108 | 123 | 1870 | 57 |
| 114 | 30 | 20 | 0 | 100 | 5 | 60 | 195 | Cu | Off | 755 | 25 | 1782 | 78 | 1173 | 35 |
| 115 | 30 | 20 | 0 | 125 | 5 | 60 | 195 | Cu | Off | 891 | 30 | 2033 | 89 | 1335 | 40 |
| 116 | 30 | 20 | 0 | 150 | 5 | 60 | 195 | Cu | Off | 1201 | 40 | 2416 | 106 | 1620 | 47 |
| 117 | 40 | 15 | 0 | 100 | 5 | 60 | 195 | Cu | Off | 560 | 19 | 1961 | 78 | 1186 | 36 |
| 118 | 40 | 15 | 0 | 125 | 5 | 60 | 195 | Cu | Off | 984 | 33 | 2348 | 96 | 1455 | 44 |

| | | | | | | | | | | | | | | | | |
|---|---|---|---|---|---|---|---|---|---|---|---|---|---|---|---|---|
| 119 | 40 | 15 | 0 | 150 | 5 | 60 | 195 | Cu | Off | 1371 | 46 | 2916 | 121 | 1827 | 55 |
| 120 | 40 | 20 | 0 | 100 | 5 | 60 | 195 | Cu | Off | 730 | 25 | 1571 | 72 | 1066 | 31 |
| 121 | 40 | 20 | 0 | 125 | 5 | 60 | 195 | Cu | Off | 952 | 32 | 1901 | 85 | 1294 | 37 |
| 122 | 40 | 20 | 0 | 150 | 5 | 60 | 195 | Cu | Off | 1124 | 38 | 2246 | 100 | 1519 | 44 |
| 123 | 40 | 20 | 10 | 100 | 5 | 60 | 195 | Cu | Off | 619 | 21 | 1643 | 71 | 1083 | 31 |
| 124 | 40 | 20 | 10 | 125 | 5 | 60 | 195 | Cu | Off | 862 | 29 | 1635 | 75 | 1142 | 33 |
| 125 | 40 | 20 | 10 | 150 | 5 | 60 | 195 | Cu | Off | 1017 | 34 | 1923 | 89 | 1351 | 38 |
| 126 | 40 | 20 | 15 | 100 | 5 | 60 | 195 | Cu | Off | 661 | 22 | 1688 | 73 | 1103 | 32 |
| 127 | 40 | 20 | 15 | 125 | 5 | 60 | 195 | Cu | Off | 834 | 28 | 1563 | 73 | 1098 | 32 |
| 128 | 40 | 20 | 15 | 150 | 5 | 60 | 195 | Cu | Off | 984 | 33 | 1792 | 83 | 1272 | 36 |
| 129 | 40 | 20 | 20 | 100 | 5 | 60 | 195 | Cu | Off | 599 | 20 | 1669 | 73 | 1080 | 32 |
| 130 | 40 | 20 | 25 | 100 | 5 | 60 | 195 | Cu | Off | 684 | 23 | 1636 | 71 | 1072 | 31 |
| 131 | 40 | 20 | 5 | 100 | 5 | 60 | 195 | Cu | Off | 707 | 24 | 1508 | 70 | 1034 | 30 |
| 132 | 40 | 20 | 5 | 125 | 5 | 60 | 195 | Cu | Off | 952 | 32 | 1751 | 80 | 1211 | 35 |
| 133 | 40 | 20 | 5 | 150 | 5 | 60 | 195 | Cu | Off | 1087 | 37 | 2047 | 94 | 1426 | 41 |
| 134 | 40 | 25 | 0 | 100 | 5 | 60 | 195 | Cu | Off | 834 | 28 | 1473 | 75 | 1119 | 32 |
| 135 | 40 | 25 | 0 | 125 | 5 | 60 | 195 | Cu | Off | 952 | 32 | 1799 | 88 | 1348 | 37 |
| 136 | 40 | 25 | 0 | 150 | 5 | 60 | 195 | Cu | Off | 1087 | 37 | 1980 | 102 | 1546 | 43 |
| 137 | 40 | 30 | 0 | 100 | 5 | 60 | 195 | Cu | Off | 1124 | 38 | 2429 | 98 | 1470 | 45 |
| 138 | 40 | 30 | 0 | 125 | 5 | 60 | 195 | Cu | Off | 1161 | 39 | 2715 | 109 | 1663 | 50 |
| 139 | 40 | 30 | 0 | 150 | 5 | 60 | 195 | Cu | Off | 1617 | 54 | 3912 | 163 | 2437 | 74 |

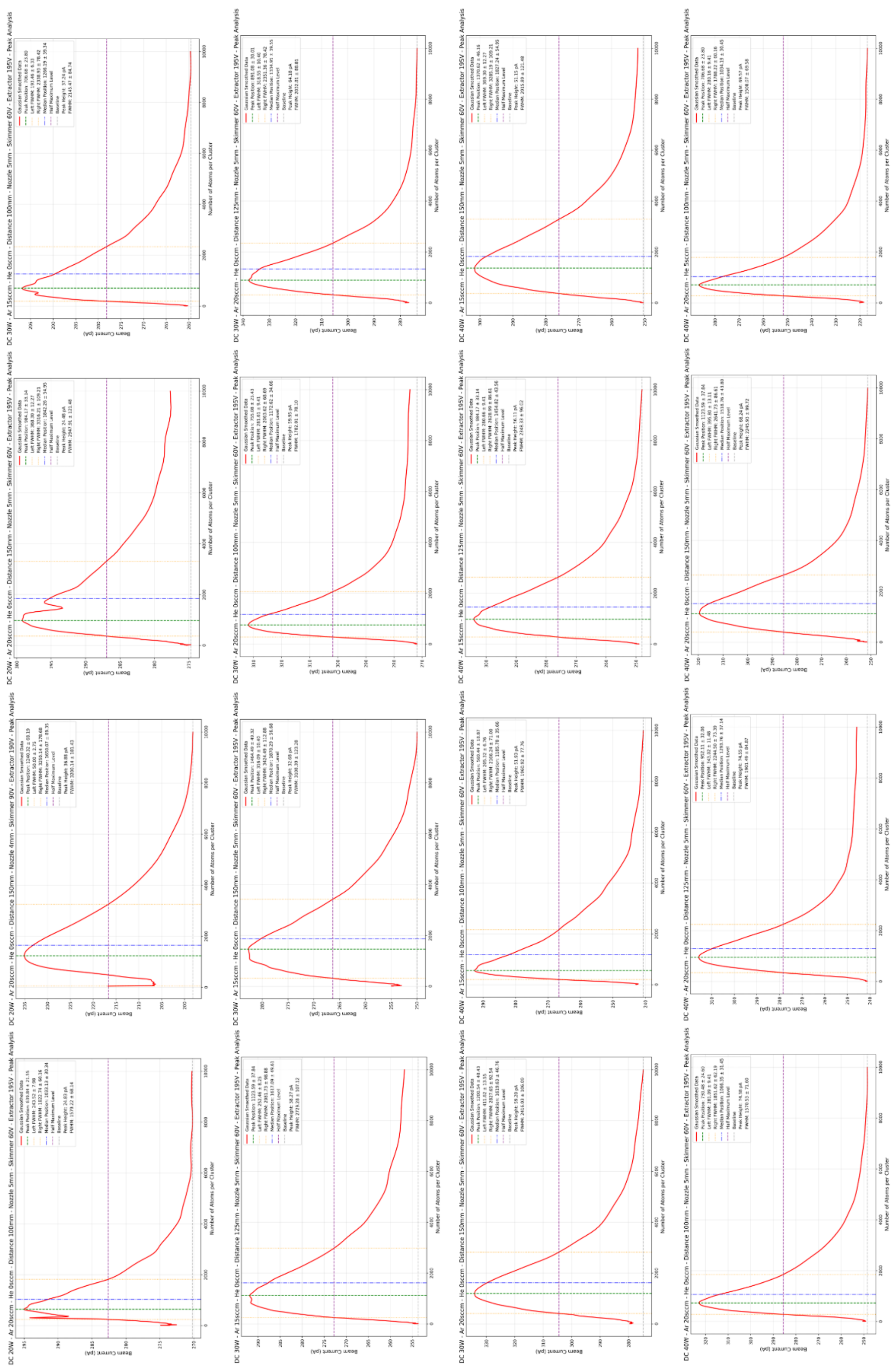

**Figure S1. Size distributions of Cu nanoparticles deposited under various deposition parameters.**

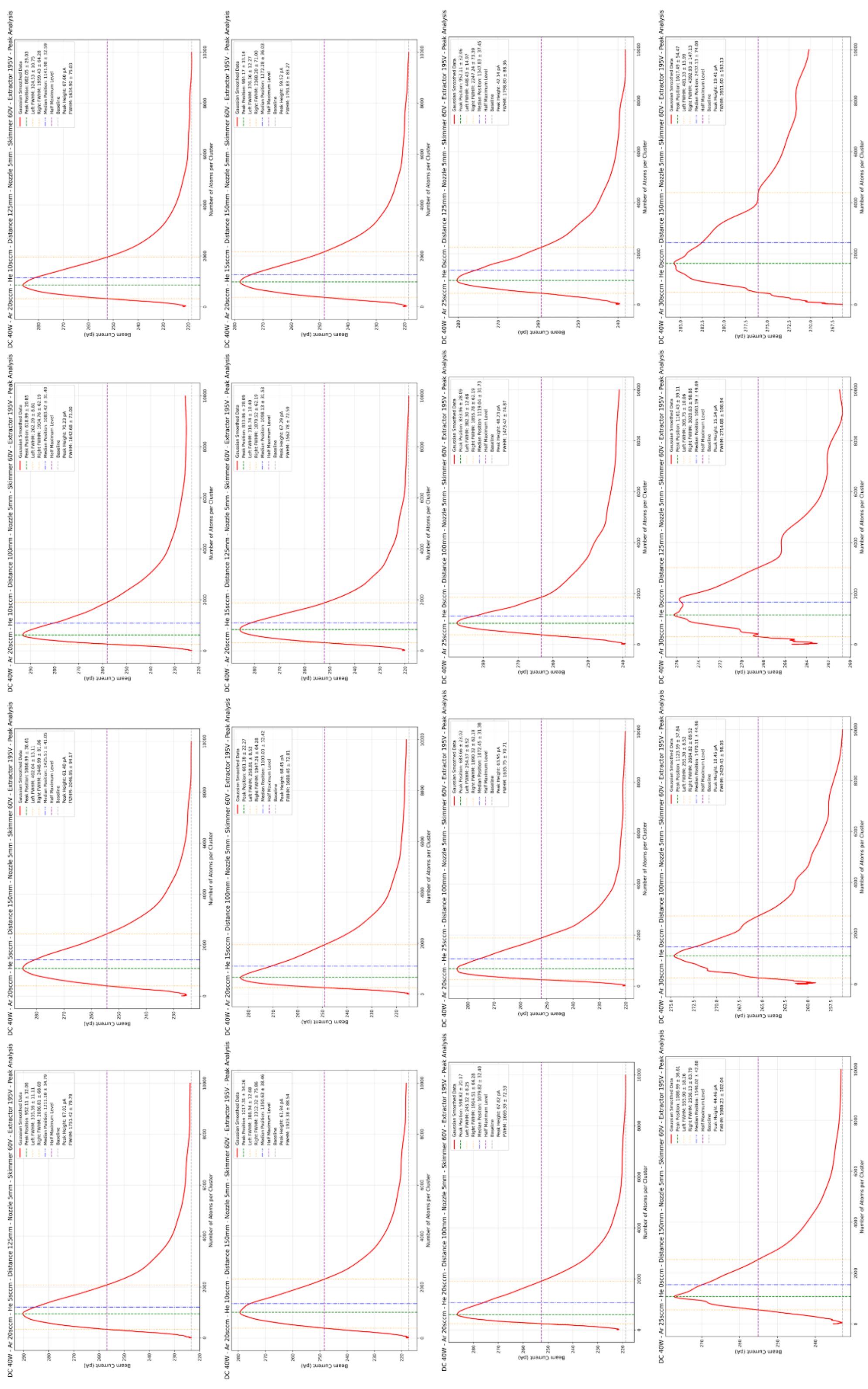

**Figure S1-continued. Size distributions of Cu nanoparticles deposited under various deposition parameters.**

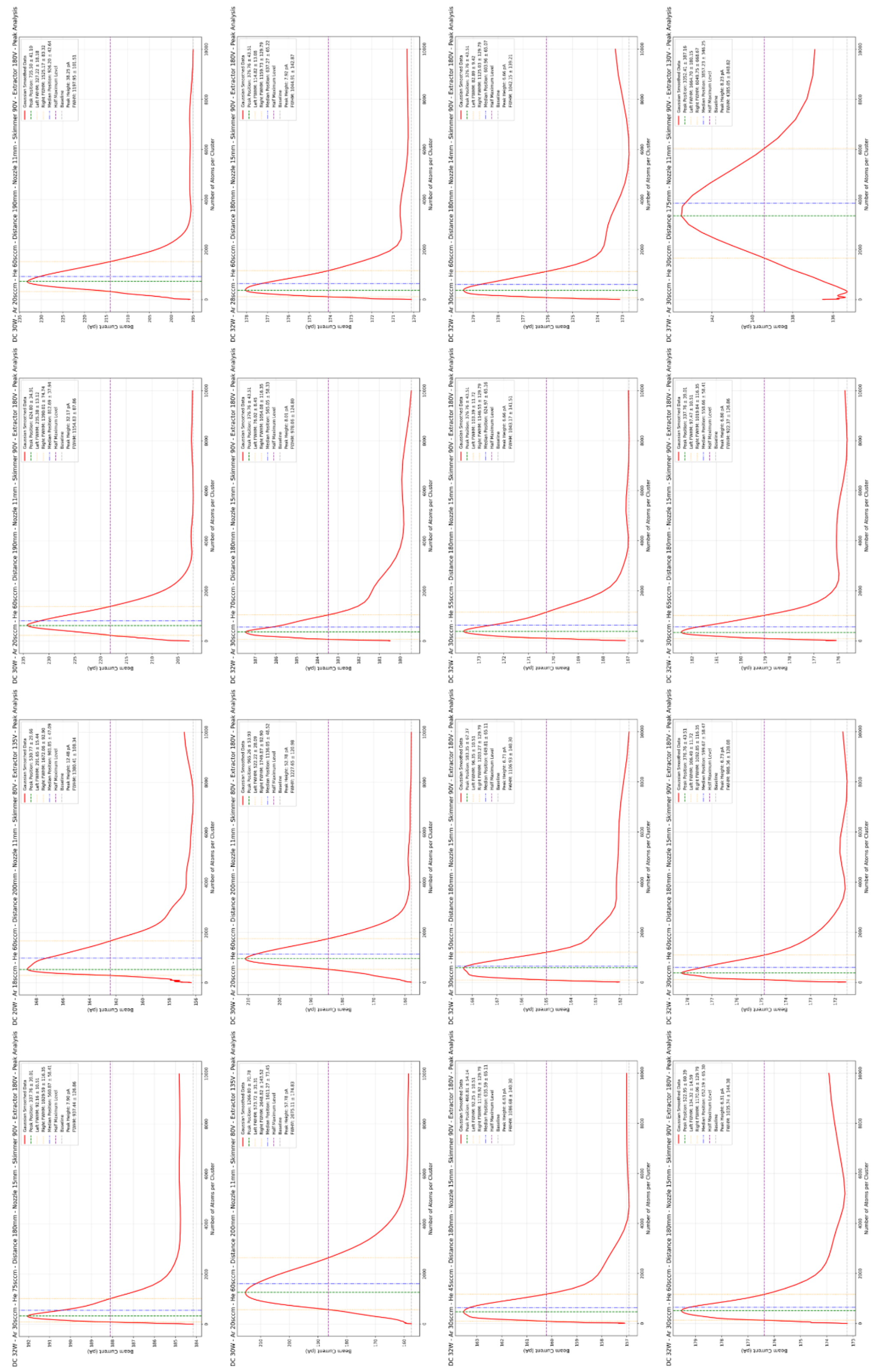

**Figure S2. Size distributions of Ta nanoparticles deposited under various deposition parameters.**

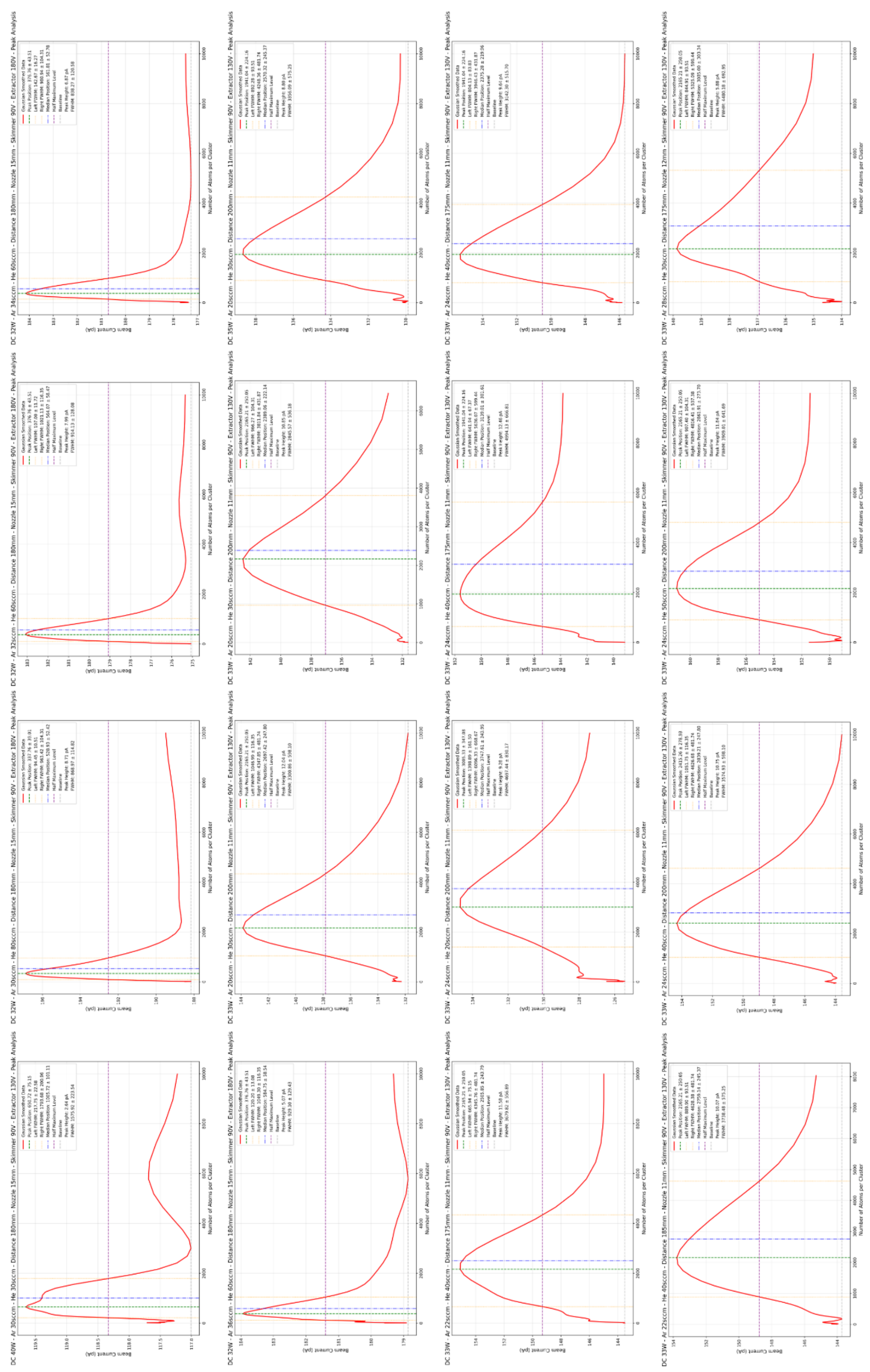

**Figure S2-continued. Size distributions of Ta nanoparticles deposited under various deposition parameters.**

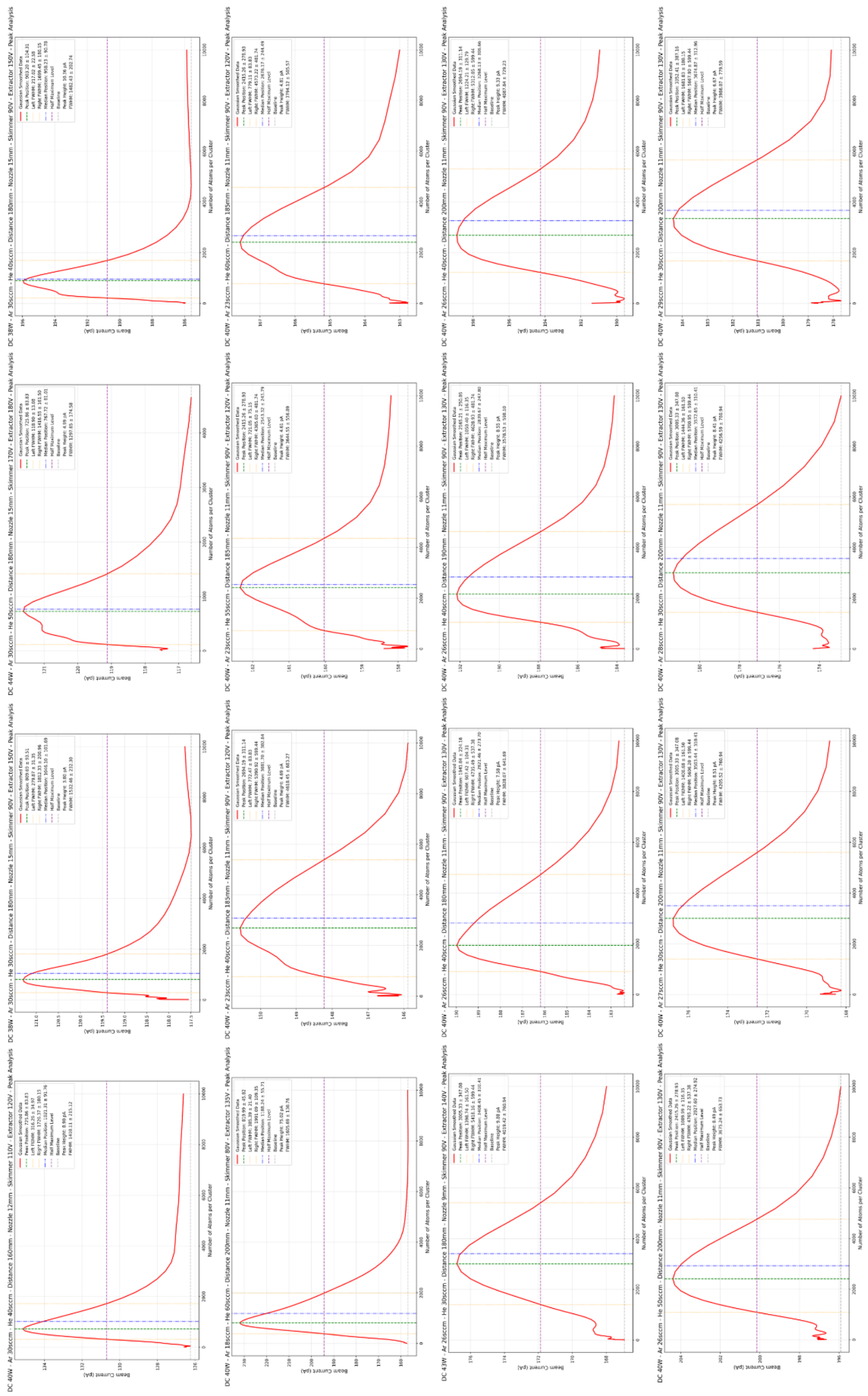

**Figure S2-continued. Size distributions of Ta nanoparticles deposited under various deposition parameters.**

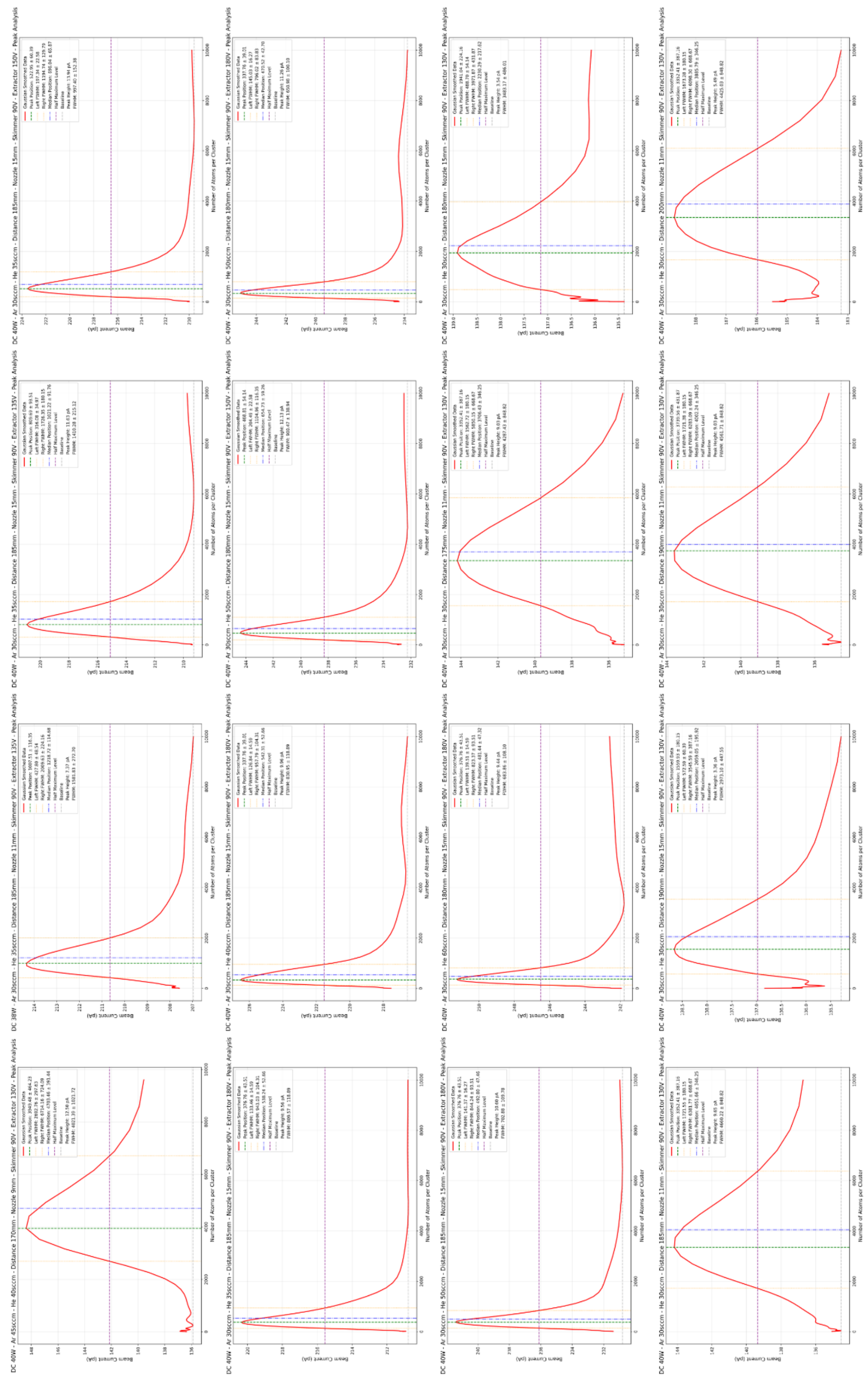

**Figure S2-continued. Size distributions of Ta nanoparticles deposited under various deposition parameters.**

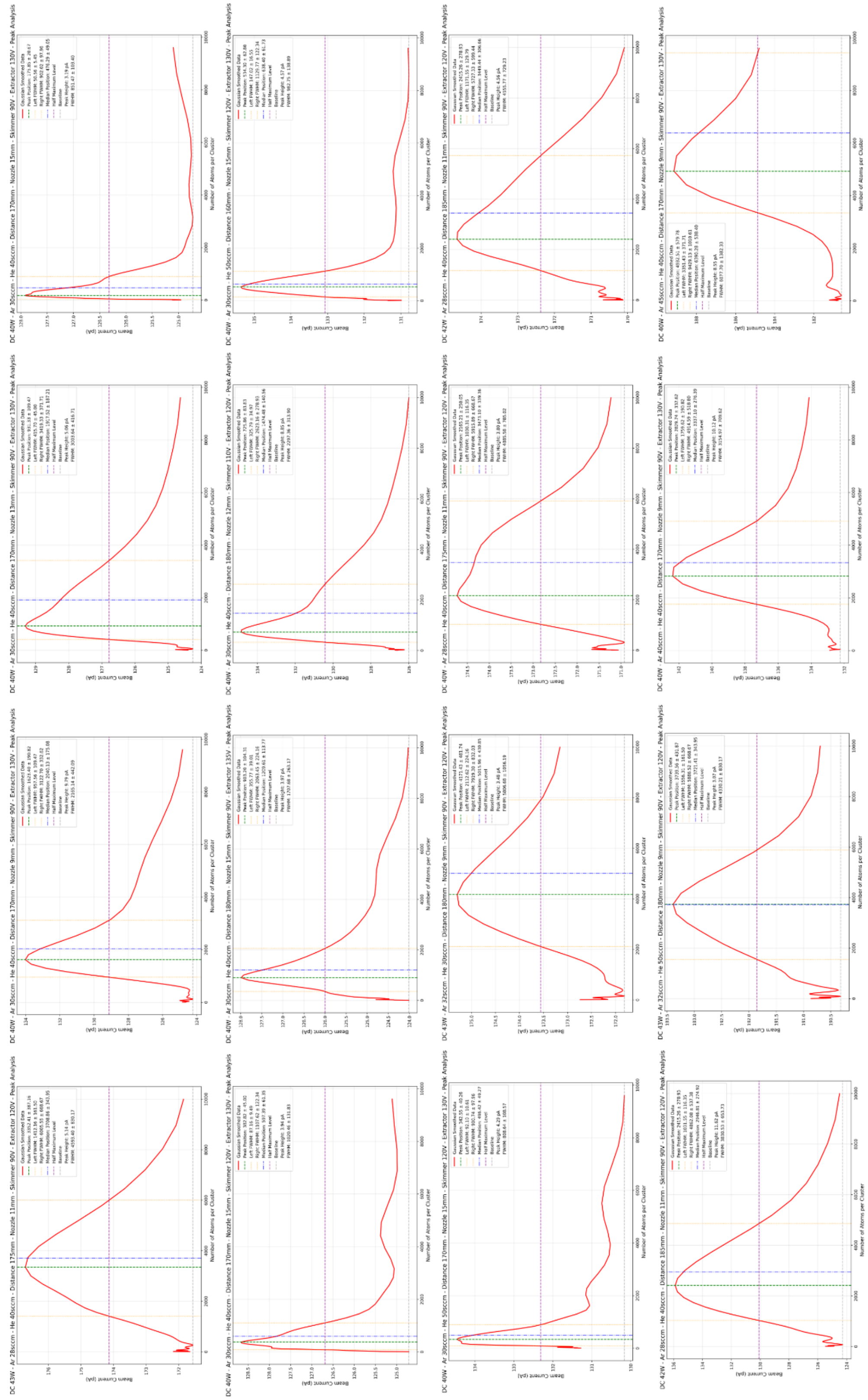

**Figure S2-continued. Size distributions of Ta nanoparticles deposited under various deposition parameters.**

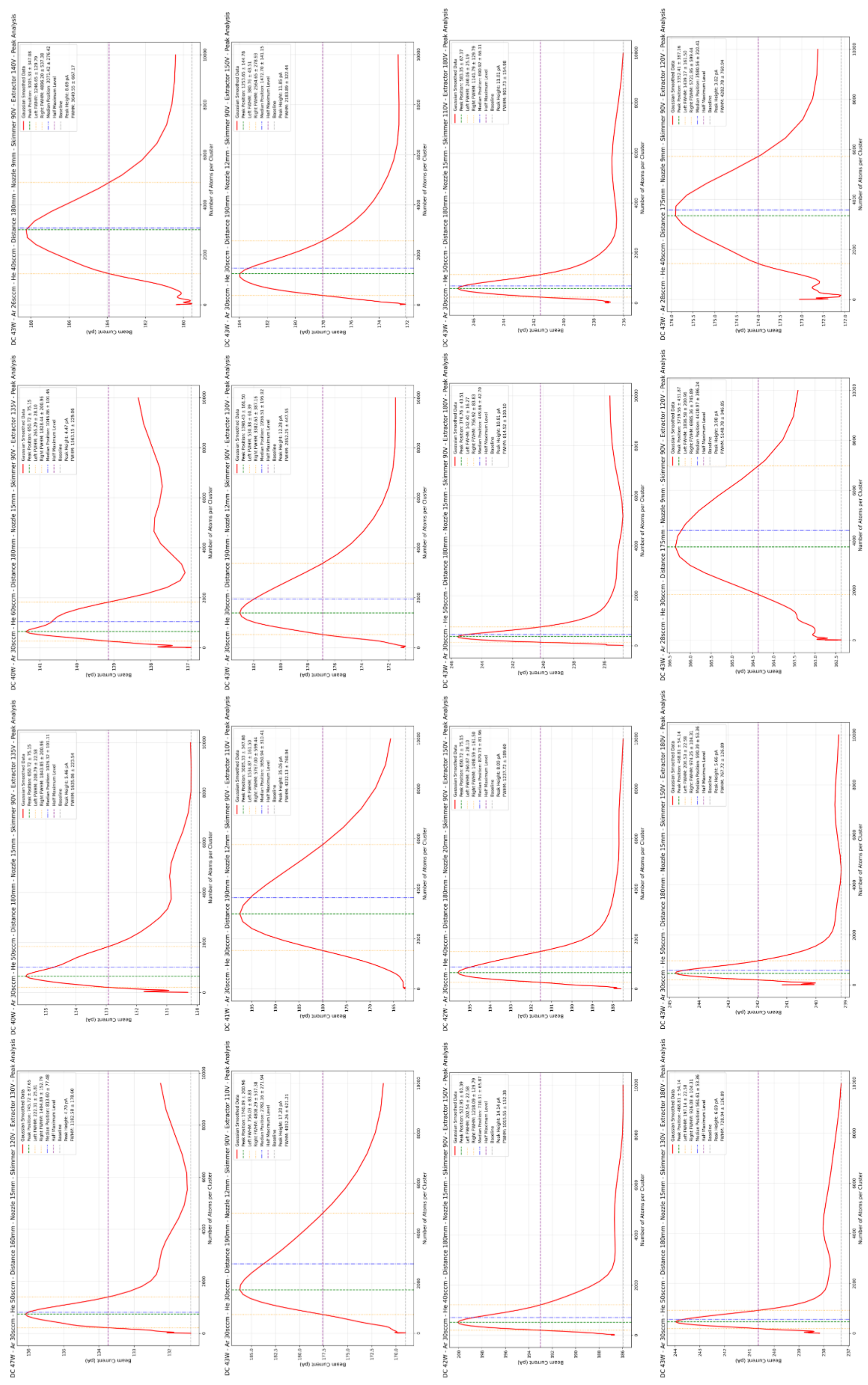

**Figure S2-continued. Size distributions of Ta nanoparticles deposited under various deposition parameters.**

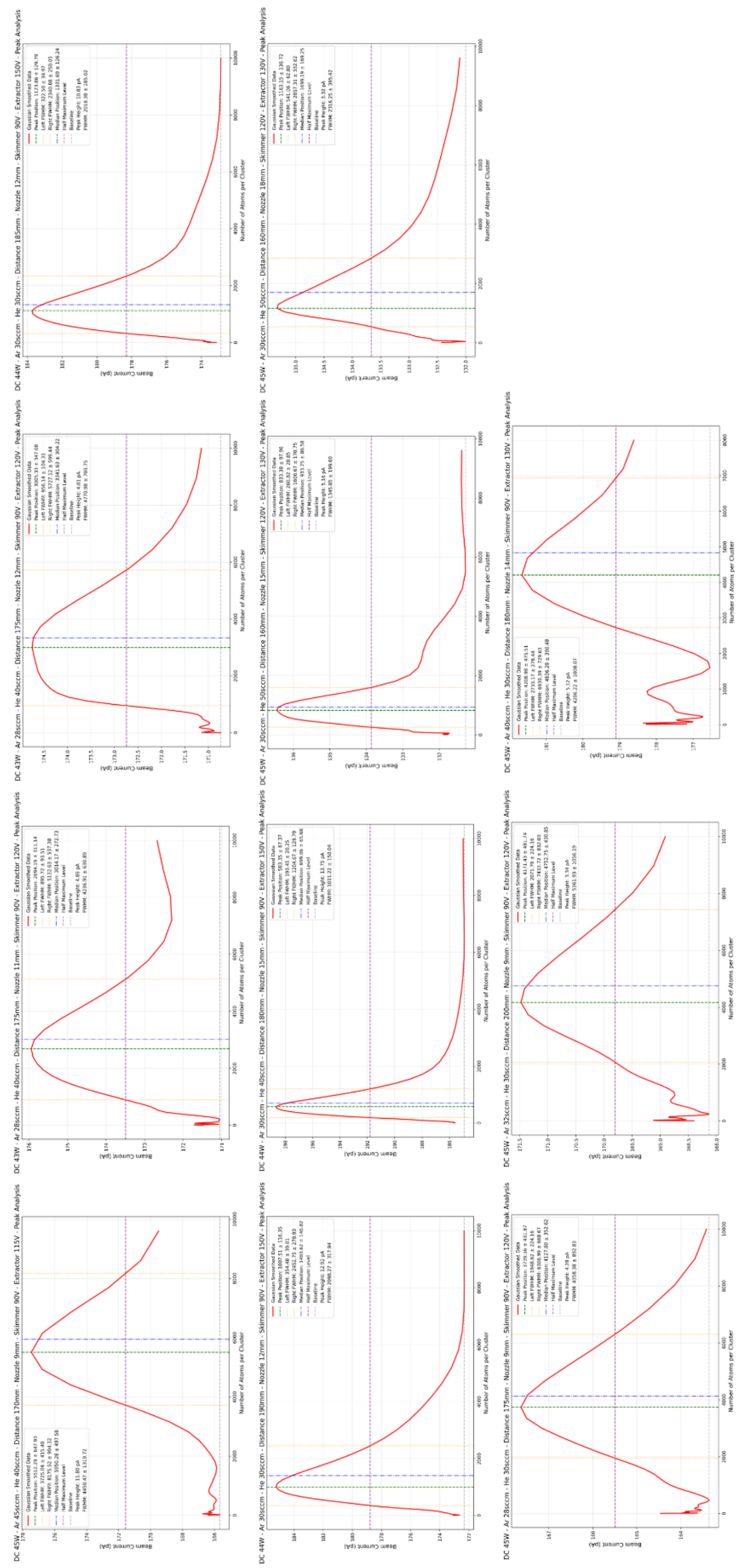

**Figure S2-continued. Size distributions of Ta nanoparticles deposited under various deposition parameters.**

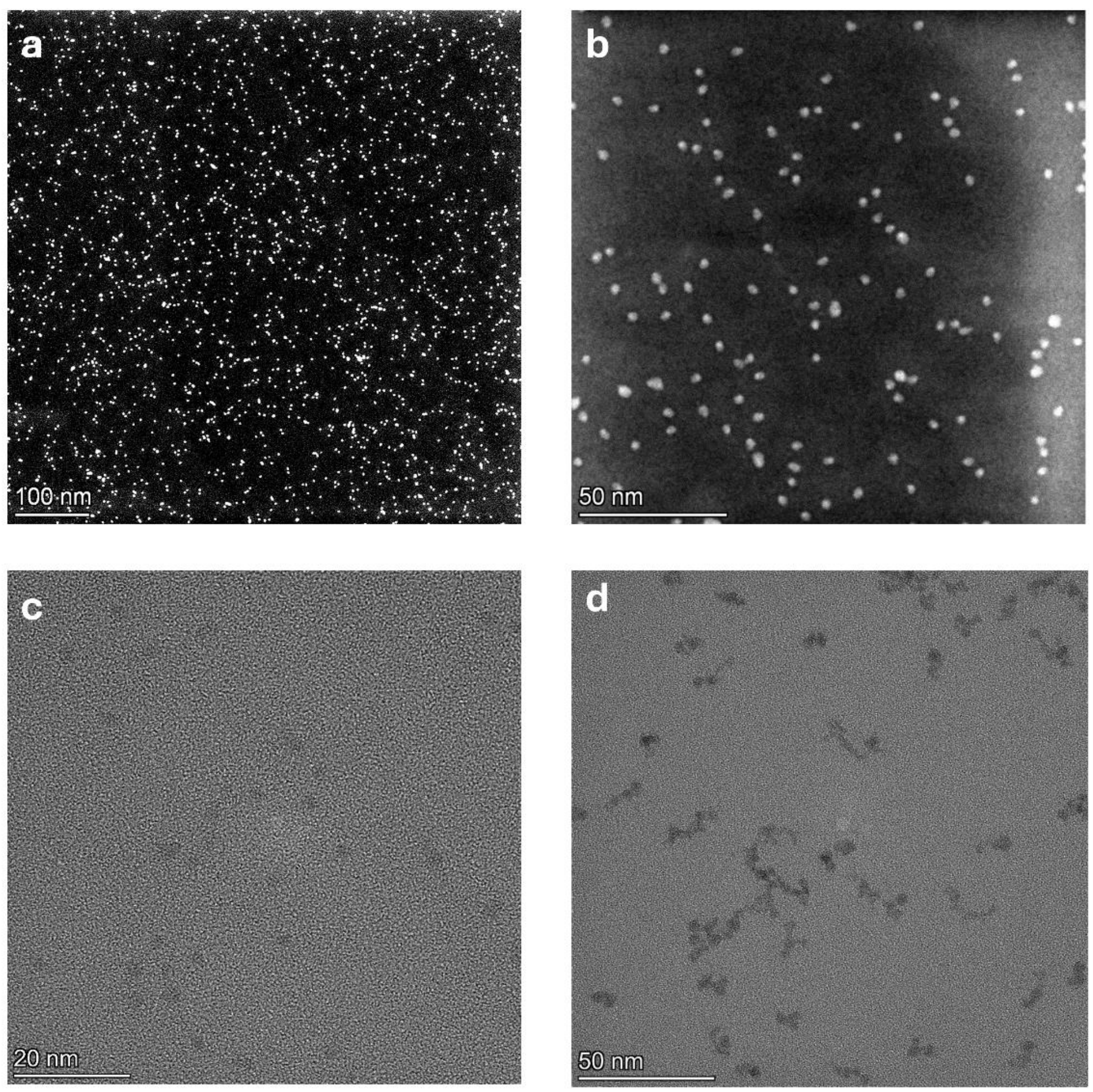

**Figure S3. High-angle annular dark-field (HAADF) and transmission electron microscopy (TEM) survey images of Cu (a-c) and Ta (d) NP deposits on Cu foil.** The survey micrographs reveal that the NPs are not always isolated and dispersed but significantly aggregated into chain-like or network structures, thus complicating the precise measurement of their sizes.